\documentclass[twocolumn,australian,english,pra]{revtex4}
\usepackage[T1]{fontenc}
\usepackage[latin9]{inputenc}
\usepackage{color}
\usepackage{amsmath}
\usepackage{amssymb}
\usepackage{stmaryrd}
\usepackage{graphicx}
\usepackage{esint}

\makeatletter
\@ifundefined{textcolor}{}
{%
 \definecolor{BLACK}{gray}{0}
 \definecolor{WHITE}{gray}{1}
 \definecolor{RED}{rgb}{1,0,0}
 \definecolor{GREEN}{rgb}{0,1,0}
 \definecolor{BLUE}{rgb}{0,0,1}
 \definecolor{CYAN}{cmyk}{1,0,0,0}
 \definecolor{MAGENTA}{cmyk}{0,1,0,0}
 \definecolor{YELLOW}{cmyk}{0,0,1,0}
}

\@ifundefined{definecolor}
 {\usepackage{color}}{}
\makeatother

\makeatother

\usepackage{babel}
\begin{document}

\title{Signifying the nonlocality of NOON states using Einstein-Podolsky-Rosen
steering inequalities}

\author{R.Y. Teh, L. Rosales-Z\'arate, B. Opanchuk and M. D. Reid}

\affiliation{Centre for Quantum and Optical Science, Swinburne University of Technology,
Melbourne, Australia}
\begin{abstract}
We construct Einstein-Podolsky-Rosen (EPR) steering signatures for
the nonlocality of the entangled superposition state described by
$\frac{1}{\sqrt{2}}\{|N\rangle|0\rangle+|0\rangle|N\rangle\}$, called
the two-mode NOON state. The signatures are a violation of an EPR
steering inequality based on an uncertainty relation. The violation
confirms an EPR steering between the two modes and involves certification
of an inter-mode correlation for number, as well as quadrature phase
amplitude measurements. We also explain how the signatures certify
an $N$th order quantum coherence, so the system (for larger $N$)
can be signified to be in a superposition of states distinct by a
mesoscopic value of the two-mode quantum number difference.  Finally,
we examine the limitations imposed for lossy scenarios, discussing
how experimental realisations may be possible for $N=2,3$.

\end{abstract}
\maketitle

\section{Introduction}

The generation and signification of a macroscopic quantum superposition
state is an outstanding challenge. Schrodinger explained that according
to quantum mechanics it is conceptually possible for a macroscopic
system (like a cat) to become entangled with a microscopic one in
such a way that a superposition of two macroscopically-distinct states
is created \cite{Schrodinger-1} . Schrodinger pointed out the paradoxical
nature of such a macroscopic system: A superposition of two macroscopically
distinguishable states cannot be interpreted as being in one or the
other of the states until measured. In realistic scenarios, couplings
to external environments make the generation of macroscopic superposition
states difficult but simpler mesoscopic realisations are feasible
\cite{ystbeamsplit-1,decoherence,cats}. 

One of the most interesting realisations is the two-mode NOON state
\cite{dowlingreveiw,NoonDowlingtheory,zeilinger nOnlocality,NOONmitchellphase,NOON_N5,noon refs exp,krusesilberhorneexp,hongoumadel,boto,appNOOn3,applications NOON,kim}:
\begin{equation}
|\psi_{\mathrm{NOON}}\rangle=\frac{1}{\sqrt{2}}\{|N\rangle|0\rangle+e^{i\phi}|0\rangle|N\rangle\}\label{eq:noon-state}
\end{equation}
Here, $N$ boson particles (or photons) are in a superposition of
being either in the first mode (denoted $a$) or the second mode (denoted
$b$). The modes may correspond to different spatial paths. Denoting
the creation and destruction operators for the two modes by $\hat{a},$
$\hat{a}^{\dagger}$ and $\hat{b}$, $\hat{b}^{\dagger}$, $|n\rangle|m\rangle$
is the eigenstate of numbers $\hat{n}_{a}=\hat{a}^{\dagger}\hat{a}$
and $\hat{n}_{b}=\hat{b}^{\dagger}\hat{b}$ with eigenvalues $n$
and $m$ respectively. Experiments have used spontaneous parametric
down conversion to generate photonic NOON states for $N$ up to 5
\cite{NOON_N5,noon refs exp,krusesilberhorneexp,zeilinger nOnlocality,hongoumadel,NOONmitchellphase,kim}.
 Recent experiments achieve Hong-Ou-Mandel interference with atoms
(for $N=2$) \cite{atom hongmandel} and proposals exist for Bose-Einstein
condensates (BEC) \cite{wallsmil}. NOON states are typically signified
by way of interference fringes or fidelity \cite{krusesilberhorneexp,noon refs exp,NOON_N5,zeilinger nOnlocality,NOONmitchellphase,applications NOON,kim}.

As $N\rightarrow\infty$, the NOON state is a superposition of two
states with macroscopically different values of quantum number $\hat{n}$
in each mode. While genuinely macroscopic systems (like a cat) would
involve many degrees of freedoms (for example many modes) \cite{cat-disc},
the NOON state (similar to other single and two-mode states studied
in the literature \cite{cats,ystbeamsplit-1}) nonetheless provides
a simple model for the Schrodinger cat paradox, as $N\rightarrow\infty$.
The NOON state superposition (\ref{eq:noon-state}) can therefore
 elucidate aspects of the transition from microscopic to macroscopic.
In order to quantify the transition, we refer to the state (\ref{eq:noon-state})
as an ``$N$-scopic superposition''.

Our motivation is to investigate the \emph{nonlocality} of the NOON
state (or of an approximate NOON state that may be generated experimentally).
While nonlocality between two microscopic systems (corresponding to
$N=1$) has been experimentally certified using Bell inequalities
\cite{loophole-1}, relatively little is known about nonlocality between
more mesoscopic systems \cite{mesobell}. In particular, it is an
important goal to experimentally verify the nonlocality of an entangled
state like that described by Schrodinger, where the system is in a
superposition of two mesoscopically distinguishable states. In this
paper, we derive a set of \emph{Einstein-Podolsky-Rosen (EPR) steering
inequalities }\cite{steer-1,ericsteer2,steerloop2} based on the number-phase
uncertainty relation 
\begin{equation}
\Delta\hat{n}\Delta\hat{P}^{N}\geq\frac{1}{2}|\langle\left[\hat{n},\hat{P}^{N}\right]\rangle|\label{eq:uncer}
\end{equation}
where $\hat{n}$ is the mode number and $\hat{P}$ is the mode quadrature
amplitude (defined below). We show how violation of these inequalities
can be used to demonstrate the nonlocality of the NOON state for arbitrary
$N$. By examining realistic scenarios for NOON states where losses
are present,  we suggest feasible tests for $N=2,3$. 

The detection of an EPR steering nonlocality between two optical systems
consisting of many photons has been experimentally verified \cite{rrmp},
but  this does not in itself imply the type of entangled state considered
by Schrodinger: By contrast, we are able to show that the violation
of the EPR steering inequalities as predicted for the NOON state certifies
the $N$-scopic nature of the entanglement of (\ref{eq:noon-state}),
which involves a superposition of number states distinct by $N$ quanta.

EPR steering has been established as a distinct type of nonlocality,
different to both Bell's nonlocality and entanglement \cite{steer-1,epr,ericsteer2}.
``Steering'' is the term used by Schrodinger \cite{schsteer} to
describe the effect where an observer at one location can apparently
change the quantum state at another $-$ the effect Einstein called
``spooky action-at-a-distance'' \cite{einsteinspooky}. Some
EPR steering and Bell inequalities have been derived for NOON states
\cite{SteeringNOON,bellNOON,onephotonnonlocality}. For $N=1$ this
led to the experimental verification of the Bell nonlocality of a
single photon \cite{onephotonnonlocality}. More recently, steering
inequalities for $N=1$ have been used to give conclusive proof of
the ``collapse of the wavefunction'' \cite{JWone photon,Furusawa}.
The EPR steering inequalities for larger $N$ may therefore open
a way to investigate such effects for a mesoscopic superposition state.

Most steering and Bell inequalities derived to date use either number
or quadrature phase amplitude measurements. The proposal of this paper
combines number and quadrature phase amplitude measurements. This
gives two advantages. First, the number measurements are useful in
optimising violation of the inequalities for entangled two-mode systems
over a range of field intensities where there is a perfect number
correlation between the modes (as with NOON states). Second, the inequalities
are based on variances and provide a simple method in nonideal scenarios
with small losses to demonstrate that the nonlocality observed in
the experiment is indeed due to a superposition of states distinct
by $\sim N$ quanta.

\textbf{\emph{Summary of paper: }}Our proposed EPR steering inequalities
are derived in Sections II and III of this paper. For $N=2$ and $\phi\neq0$
we show that a suitable signature for the steering nonlocality of
a NOON state is the violation of the EPR steering inequality
\begin{equation}
\Delta_{inf}\hat{n}_{b}\Delta_{inf}(\hat{P}_{b}^{2})\geq|\langle\hat{C}_{b}\rangle|_{inf}/2\label{eq:steerinequintro}
\end{equation}
where $\hat{C}_{b}=2\hat{X}_{b,\pi/4}-\hat{X}_{b}^{2}-\hat{P}_{b}^{2}$.
Here we define the rotated quadrature phase amplitudes for mode $b$
as $\hat{X_{b}}_{,\theta}=\hat{X}_{b}\cos\theta+\hat{P}_{b}\sin\theta$
and $\hat{P}_{b,\theta}=-\hat{X}_{b}\sin\theta+\hat{P}_{b}\cos\theta$
where $\hat{X}_{b}=\hat{b}+\hat{b}^{\dagger}$, $\hat{P}_{b}=(\hat{b}-\hat{b}^{\dagger})/i$.
Also $\hat{X}_{a}=\hat{a}+\hat{a}^{\dagger}$, $\hat{P}_{a}=(\hat{a}-\hat{a}^{\dagger})/i$.
The $\Delta_{inf}\hat{n}_{b}$ is the uncertainty in the prediction
for $\hat{n}_{b}$ based on measurement of $\hat{n}_{a}$. Similarly
$\Delta_{inf}\hat{P}_{b}^{2}$ is the uncertainty in $\hat{P}_{b}^{2}$
based on the measurement $\hat{X}_{a}$; and $|\langle\hat{C}_{b}\rangle|_{inf}$
is the magnitude of the mean value of $\hat{C}_{b}$ based on the
measurement $\hat{X}_{a}$. An EPR steering inequality is obtained
by replacing the quantities of an uncertainty relation (in this case
(\ref{eq:uncer})) with their predicted (``inferred'') values \cite{eprr-2,rrmp,mdrspin}.
In Section II, we summarise the Local Hidden State (LHS) Model developed
by Wiseman, Jones and Doherty \cite{steer-1}. Using the methods of
Cavalcanti et al \cite{ericsteer2}, we prove that (\ref{eq:steerinequintro})
is a steering inequality the violation of which falsifies LHS models,
so that steering of the mode $b$ (by measurements on the mode $a$)
can be confirmed. 

In Sections II and III, we provide similar inequalities for arbitrary
$N$, including one for odd $N$ and $\phi=0$. Specifically, EPR
steering of the mode $b$ is confirmed if 
\begin{equation}
E_{N}^{(p)}=\frac{\Delta_{inf}\hat{n}_{b}\Delta_{inf}\hat{P}_{b}^{N}}{\frac{1}{2}|\langle\left[\hat{n}_{b},\hat{P^{N}}\right]\rangle|_{inf}}<1\label{eq:steerineq1p-1}
\end{equation}
or 
\begin{equation}
E_{N}^{(x)}=\frac{\Delta_{inf}\hat{n}_{b}\Delta_{inf}\hat{X}_{b}^{N}}{\frac{1}{2}|\langle\left[\hat{n}_{b},\hat{X}_{b}^{N}\right]\rangle|_{inf}}<1\label{eq:steerineq2x-1}
\end{equation}
 For the ideal NOON state, $\Delta_{inf}\hat{n}_{b}=0$ and the usefulness
of the inequality depends on whether the denominator is nonzero. We
show that the first criterion is useful provided $\cos\phi\neq0$
for $N$ odd or $\sin\phi\neq0$ for $N$ even, and the second criterion
is useful for all $N$ provided $\sin\phi\neq0$. We explain in Section
VII for $N$ up to $3$ how the denominator of the inequality can
be measured via homodyne detection. For $N=1$ the inequality is becomes
straightforwardly 
\begin{equation}
\Delta_{inf}\hat{n}_{b}\Delta_{inf}\hat{P}_{b}<|\langle\hat{X}_{b}\rangle|_{inf}/2\label{eq:steeringcritP1-1}
\end{equation}
The cases of $N=1$ and $N=2$ are analysed in detail in Sections
IV and V. The explanation of how the steering inequalities signify
an $N$-scopic superposition state is given in Section VIII.

The inequalities (\ref{eq:steerinequintro}-\ref{eq:steerineq2x-1})
involve measurement of number ($\hat{n}_{a}$, $\hat{n}_{b}$) and
hence have the drawback of low detection efficiencies (in the photonic
case). In the first instance, we propose that the correlation be
established by postselection of the events where a total of $N$ quanta
(photons) are detected at the sites of both modes.  A second problem
is distinguishing between the detection of two and one photons at
a given site. Here, beam splitters or photon number-resolving detectors
could be used \cite{krusesilberhorneexp,NOON_N5} in conjunction with
postselection over events where a total $N$ photons is counted.
The measurement of observables $\hat{X}^{N}$, $\hat{P}^{N}$ is achieved
via optical homodyne techniques that are highly efficient. Nonetheless,
we explain in Sections VI, VII and IX that losses have a significant
effect (measurement efficiencies of $\eta>0.94$ are required for
$N=3$) and that care needs to be taken to avoid possible loopholes
created by asymmetrical losses for the number and quadrature measurements.

\section{EPR Steering inequalities based on uncertainty relations}

In this Section we give the formal derivation of the EPR steering
criteria summarised in the Introduction I. We show in Section III
how we can use the inequalities to detect EPR steering for a NOON
state.

\subsection{EPR steering inequalities}

EPR steering is verified as a failure of Local Hidden State models
(LHS). The LHS model was pioneered in the papers by Wiseman, Jones
and Doherty \cite{steer-1} and is based on the Local Hidden Variable
models considered by Bell \cite{Bell-1}. We define two subsystems
$A$ and $B$ and consider space-like separated measurements on each
of them. The measurements are described quantum mechanically by observables
$\hat{X}_{A}(\theta)$ and $\hat{X}_{B}(\phi)$ (respectively) and
the outcomes are given by the numbers $X_{A}(\theta)$ and $X_{B}(\phi)$
(written without the ``hats''). Here $\theta$ and $\phi$ denote
the measurement choice at the locations $A$ and $B$. To prove Bell's
nonlocality, one falsifies a description of the statistics based on
a Local Hidden Variable model, where the averages are given as
\begin{eqnarray}
\langle X_{B}(\phi)X_{A}(\theta)\rangle & = & \int_{\lambda}d\lambda P(\lambda)\langle X_{B}(\phi)\rangle_{\lambda}\langle X_{A}(\theta)\rangle_{\lambda}\nonumber \\
\label{eq:lhv}
\end{eqnarray}
Here $\int_{\lambda}P(\lambda)d\lambda=1$ so that the $P(\lambda)$
is a probability density (or probability if the integral is replaced
by a discrete summation, as explained in Bell's papers \cite{Bell-1}).
The $\lambda$ denotes a set of variables $\{\lambda\}$ that take
the role of the hidden variables as postulated in Bell's model. The
$\langle X_{A}\rangle_{\lambda}$ denotes the average of the results
$X_{A}$ for the system in the particular hidden variable state denoted
by $\lambda$; and similarly for $\langle X_{B}\rangle_{\lambda}$.
The $P(\lambda)$ is independent of the $\theta$ and $\phi$. The
factorisation that occurs for the moments in the integrand is due
to the assumption of ``locality'' \cite{Bell-1}. 

To prove EPR steering of subsystem $B$, we need to falsify a description
of the statistics based on a Local Hidden State (LHS) model where
the averages are given as \cite{steer-1,ericsteer2}
\begin{eqnarray}
\langle X_{B}(\phi)X_{A}(\theta)\rangle & = & \int_{\lambda}d\lambda P(\lambda)\langle X_{B}(\phi)\rangle_{\lambda,\rho}\langle X_{A}(\theta)\rangle_{\lambda}\nonumber \\
\label{eq:lhs}
\end{eqnarray}
Here an extra condition is placed on the average $\langle X_{B}\rangle_{\lambda}$.
The $\rho$ subscript denotes that the average is to be consistent
with that of a \emph{quantum density operator} $\rho_{\lambda}^{B}$.
This is the case for all choices $\phi$ of measurement at $B$. For
example, if $X_{B}(\theta)=X_{B}$ and $X_{B}(\pi/2)=P_{B}$ then
the statistics for the LHS model must be consistent with a local uncertainty
principle namely $\langle(X_{B}-\langle X_{B}\rangle)^{2}\rangle_{\lambda}\langle(P_{B}-\langle P_{B}\rangle)^{2}\rangle_{\lambda}\geq1$.
The $\rho_{\lambda}^{B}$ is an example of a Local Quantum State (for
site $B$). No such constraint is made for the moments $\langle X_{A}(\theta)\rangle_{\lambda}$,
written without the subscript.

In this paper we consider three quantum observables defined through
the uncertainty relation: 
\begin{equation}
\Delta\hat{\sigma}_{B}^{X}\Delta\hat{\sigma}_{B}^{Y}\geq|\langle\hat{\sigma}_{B}^{Z}\rangle|/2\label{eq:unc}
\end{equation}
Following the approach given in Refs. \cite{rrmp,eprr-2} used to
derive a criterion for the EPR paradox \cite{epr} and also for EPR
steering \cite{steer-1,rrmp}, we consider the average conditional
uncertainty $\Delta_{inf}\sigma_{B}^{X}$ defined by 
\begin{equation}
(\Delta_{inf}\hat{\sigma}_{B}^{X})^{2}=\sum_{x_{j}^{A}}P(x_{j}^{A})(\Delta(\sigma_{B}^{X}|x_{j}^{A}))^{2}\label{eq:inf-2}
\end{equation}
Here, we denote the possible results of a measurement $\hat{X}_{A}$
at $A$ by $\{x_{j}^{A}\}$. $P(x_{j}^{A})$ is the probability for
obtaining the result $x_{j}^{A}$. The uncertainty (\ref{eq:inf-2})
is a measure of the (average) uncertainty in the \emph{inferred} value
(which we take to be the mean of the conditional distribution $P(\sigma_{B}^{X}|x_{j}^{A})$)
for a measurement $\hat{\sigma}_{B}^{X}$ at $B$ given a measurement
$\hat{X}_{A}$ at $A$. Specifically, $(\Delta(\sigma_{B}^{X}|x_{j}^{A}))^{2}$
is the variance of the conditional distribution $P(\sigma_{B}^{X}|x_{j}^{A})$.
We define similarly
\begin{equation}
(\Delta_{inf}\hat{\sigma}_{B}^{Y})^{2}=\sum_{y_{j}^{A}}P(y_{j}^{A})(\Delta(\sigma_{B}^{X}|y_{j}^{A}))^{2}\label{eq:defnycond}
\end{equation}
noting that the $\left\{ y_{j}\right\} $ is the set of results for
a measurement $\hat{Y}_{A}$ made at $A$ to infer the value of the
measurement of $\hat{\sigma}_{B}^{Y}$ at $B$. Further, we define
an (average) inferred value for the modulus of the mean of measurement
of $\hat{\sigma}_{B}^{Z}$ given a measurement $\hat{Z}_{A}$ at $A$
as 
\begin{equation}
|\langle\hat{\sigma}_{B}^{Z}\rangle|_{inf}=\sum_{z_{j}^{A}}P(z_{j}^{A})|\langle\sigma_{B}^{Z}\rangle_{z_{j}^{A}}|\label{eq:defnzcond}
\end{equation}
Here $\langle\sigma_{B}^{Z}\rangle_{z_{j}^{A}}$ is the mean of the
conditional distribution $P(\sigma_{B}^{Z}|z_{j}^{A})$ and the $\{z_{j}\}$
is the set of values for a measurement $\hat{Z}_{A}$ at $A$, that
we use to infer outcomes for $\hat{\sigma}_{B}^{Z}$. Using these
definitions, we can prove the following result \cite{mdrspin}.

\textbf{\emph{Result (1): $-$ The EPR steering inequality }}

The LHS model (\ref{eq:lhs}) implies the inequality 
\begin{eqnarray}
(\Delta_{inf}\hat{\sigma}_{B}^{X})(\Delta_{inf}\hat{\sigma}_{B}^{Y}) & \geq & |\langle\hat{\sigma}_{B}^{Z}\rangle|_{inf}/2\label{eq:steerthree}
\end{eqnarray}
Hence, violation of this inequality (called an EPR steering inequality)
implies failure of the LHS model (Eq. (\ref{eq:lhs})), and therefore
steering of system $B$ by (measurements at $A$). The proof is given
in  the Appendix \ref{sec:AppendixProofs}.

\section{Steering inequalities for the NOON state }

To arrive at a steering signature for a NOON state, we consider the
three observables for each mode: number $\hat{n}$, and the two quadrature
phase amplitudes $\hat{X}$ and $\hat{P}$. Specifically: $\hat{n}_{a}=\hat{a}^{\dagger}\hat{a}$,
$\hat{X}_{a}=\hat{a}+\hat{a}^{\dagger}$ and $\hat{P_{a}}=(\hat{a}-\hat{a}^{\dagger})/i$,
and $\hat{n}_{b}=\hat{b}^{\dagger}\hat{b}$, $\hat{X}_{b}=\hat{b}+\hat{b}^{\dagger}$
and $\hat{P}_{b}=(\hat{b}-\hat{b}^{\dagger})/i$. \textcolor{black}{Where
the notation is clear, we omit the ``hat'' for these operators.
Using the Result (1) given by Eq. (\ref{eq:steerthree}), we can write
down EPR steering criteria associated with the three observables:
We certify EPR steering (of $B$ by $A$) if either one of the following
hold:}

\textcolor{black}{
\begin{eqnarray}
\Delta_{inf}n_{b}\Delta_{inf}\left(P_{b}^{N}\right) & < & |\langle\left[n_{b},P_{b}^{N}\right]\rangle|_{inf}/2\label{eq:steerineqNP-1-1}
\end{eqnarray}
}and 

\textcolor{black}{
\begin{eqnarray}
\Delta_{inf}n_{b}\Delta_{inf}(X_{b}^{N}) & < & |\langle\left[n_{b},X_{b}^{N}\right]\rangle|_{inf}/2\label{eq:steer_x^N-2}
\end{eqnarray}
}Here, $\Delta_{inf}n_{b}$ refers to the average uncertainty of the
result for $n_{b}$ given a measurement $\hat{O}_{n}$ at $A$, as
defined by (\ref{eq:inf-2}). Similarly, $\Delta_{inf}P_{b}^{N}$
refers to the average uncertainty of the result for $P_{b}^{N}$ given
a measurement $\hat{O}_{p}$ at $A$. The $\Delta_{inf}X_{b}^{N}$
refers to the average uncertainty of the result for $X_{b}^{N}$ given
a measurement $\hat{O}_{x}$ at $A$. The $|\langle\hat{C}\rangle|_{inf}$
where $\hat{C}=\left[n_{b},P_{b}^{N}\right]$ (or $\left[n_{b},X_{b}^{N}\right]$)
is defined similarly, by (\ref{eq:defnzcond}), as the average value
of the modulus of the expectation value of $\hat{C}$ conditioned
on a measurement $\hat{O}_{c}$ at $A$. The steering inequalities
of this paper take $\hat{O}_{n}=\hat{n}_{a}$, $\hat{O}_{p}=\hat{X}_{a}$,
$\hat{O}_{x}=\hat{X}_{a}$ and $\hat{O}_{c}=\hat{X}_{a}$. The motivation
for this choice is explained in Section IV. 

To evaluate the right side of the inequalities (\ref{eq:steerineqNP-1-1}-\ref{eq:steer_x^N-2}),
we determine the commutation relations:  $[n,X]=-iP$ and $[n,P]=iX$.
\textcolor{black}{By ordering the $P$'s to be always on the left
of the $X$'s and }since $[X,P]=2i$\textcolor{black}{, we arrive
at the commutation relation $\left[X,P^{k}\right]=2ikP^{k-1}$. It
can be shown that $\left[n,P^{N}\right]=iN\{P^{N-2}\left[PX+\left(N-1\right)i\right]\}$}
and \textcolor{black}{$\left[n,X^{N}\right]=-iN\{X^{N-2}\left[XP-\left(N-1\right)i\right]\}$.
We use this result to further evaluate the right side of the steering
inequalities. Most generally, the right side of the steering inequality
}(\ref{eq:steerineqNP-1-1})\textcolor{black}{{} can be written 
\begin{eqnarray}
|\langle\left[n_{b},P_{b}^{N}\right]\rangle|_{inf} & = & N|\langle P_{b}^{N-1}X_{b}+i(N-1)P_{b}^{N-2}\rangle|_{inf}\nonumber \\
\label{eq:C}
\end{eqnarray}
so that the procedure is to measure the modulus of the expectation
value of the measurement $\hat{C}=P_{b}^{N-1}X_{b}+i(N-1)P_{b}^{N-2}$
made on mode $b$, given a specific result for a measurement $\hat{O}_{c}$
is made on mode $a$, and then take the weighted average as defined
by (\ref{eq:defnzcond}). We discuss methods for measuring $P_{b}^{N-1}X_{b}$
where $N=1,2,3$ in Section VII below. }

\textcolor{black}{To investigate whether the steering inequalities
will be useful for the NOON states (\ref{eq:noon-state}) with phase
$\phi$,}\textcolor{green}{{} }\textcolor{red}{{} }\textcolor{black}{we
evaluate the prediction for the right side of the steering inequality
(\ref{eq:steerineqNP-1-1}) in the general NOON case. We will take
$\hat{O}_{c}$ to be the measurement $X_{a}$ and denote the result
of that measurement by $x$.  We find
\begin{eqnarray}
|\langle\left[n_{b},P_{b}^{N}\right]\rangle|_{inf} & = & N|\langle P_{b}^{N-2}(P_{b}X_{b}+\left(N-1\right)i\rangle|_{inf}\nonumber \\
 & = & N|(\langle b^{N}\rangle+(-1)^{N+1}\langle b^{\dagger N}\rangle|_{inf}\nonumber \\
 & = & N\sqrt{N!}|\frac{\left[e^{i\phi}+\left(-1\right)^{N+1}e^{-i\phi}\right]}{2}|\nonumber \\
 &  & \times\intop_{-\infty}^{\infty}|\langle x|0\rangle\langle x|N\rangle|\,dx\label{eq:eqncal-1}
\end{eqnarray}
}where $|x\rangle$ are the eigenstates of $X$. The cases $N=1$
and $N=2$ are presented in the Sections IV and V below. We find similarly\textcolor{red}{}\textcolor{blue}{}

\textcolor{black}{
\begin{eqnarray}
|\langle\left[n_{b},X_{b}^{N}\right]\rangle|_{inf} & = & N|\left\langle X_{b}^{N-2}\left[X_{b}P_{b}-\left(N-1\right)i\right]\right\rangle |_{inf}\nonumber \\
 & = & N|-\langle b^{N}\rangle+\langle\left(b^{\dagger}\right)^{N}\rangle|_{inf}\nonumber \\
 & = & N\sqrt{N!}|\sin\phi|\intop_{-\infty}^{\infty}|\langle x|0\rangle\langle x|N\rangle|\,dx\nonumber \\
\label{eq:eqncal2}
\end{eqnarray}
}Now we determine when each of the steering criteria \textcolor{black}{(\ref{eq:steerineqNP-1-1})}
and \textcolor{black}{(\ref{eq:steer_x^N-2})} will be useful. For
the NOON state, the mode numbers are always correlated, and we observe
that $\Delta_{inf}n_{b}=0$. Hence either of the steering criteria
\textcolor{black}{(\ref{eq:steerineqNP-1-1})} and \textcolor{black}{(\ref{eq:steer_x^N-2})}
will be effective to detect steering in NOON states, \emph{provided}
that the right side of the inequality is not zero, and provided the
variances $\Delta_{inf}(X_{b}^{N})$, $\Delta_{inf}(P_{b}^{N})$ are
finite. Since the integral $\intop_{-\infty}^{\infty}|\langle x|0\rangle\langle x|N\rangle|\,dx$
is nonzero for the NOON state, we see from the expressions (\ref{eq:eqncal-1})
and (\ref{eq:eqncal2}) that the condition for the right side of the
inequalities\textcolor{black}{{} (\ref{eq:steerineqNP-1-1})} and \textcolor{black}{(\ref{eq:steer_x^N-2})}
to be nonzero is: for $N$ odd, $\cos\phi\neq0$ and $\sin\phi\neq0$
respectively; for $N$ even, $\sin\phi\neq0$ in both cases.

To summarise, we rewrite the EPR steering criteria \textcolor{black}{(\ref{eq:steerineqNP-1-1})}
and \textcolor{black}{(\ref{eq:steer_x^N-2})} as 
\begin{equation}
E_{N}^{(p)}=\frac{\Delta_{inf}n_{b}\Delta_{inf}P_{b}^{N}}{\frac{1}{2}|\langle\left[n_{b},P_{b}^{N}\right]\rangle|_{inf}}<1\label{eq:steerineq1p}
\end{equation}
and 
\begin{equation}
E_{N}^{(x)}=\frac{\Delta_{inf}n_{b}\Delta_{inf}X_{b}^{N}}{\frac{1}{2}|\langle\left[n_{b},X_{b}^{N}\right]\rangle|_{inf}}<1\label{eq:steerineq2x}
\end{equation}
 Steering is obtained if $E_{N}^{(x/p)}<1$. Either criterion is sufficient
to certify an EPR paradox, or EPR steering. For the NOON state $|\psi_{\mathrm{NOON}}\rangle=\frac{1}{\sqrt{2}}\{|N\rangle|0\rangle+e^{i\phi}|0\rangle|N\rangle\}$
the first criterion is useful provided $\cos\phi\neq0$ for $N$ odd
or $\sin\phi\neq0$ for $N$ even, and the second criterion is useful
for all $N$ provided $\sin\phi\neq0$.\textcolor{red}{}\textcolor{black}{{}
We comment that the right side of the steering inequalities (\ref{eq:steerineqNP-1-1})}
and \textcolor{black}{(\ref{eq:steer_x^N-2}) needs to be }\textcolor{black}{\emph{measured}}\textcolor{black}{{}
in the experiment. We examine how this can be done below in Section
VII, finding that cases of low $N$ are much more accessible to experiment.
We also point out that except where $N=1$ or $2$, the equivalence
of the first two lines in equations (\ref{eq:eqncal-1}) and (\ref{eq:eqncal2})
holds only for the expectation values as calculated for the ideal
NOON state (\ref{eq:noon-state}).} 
\begin{figure}

\includegraphics{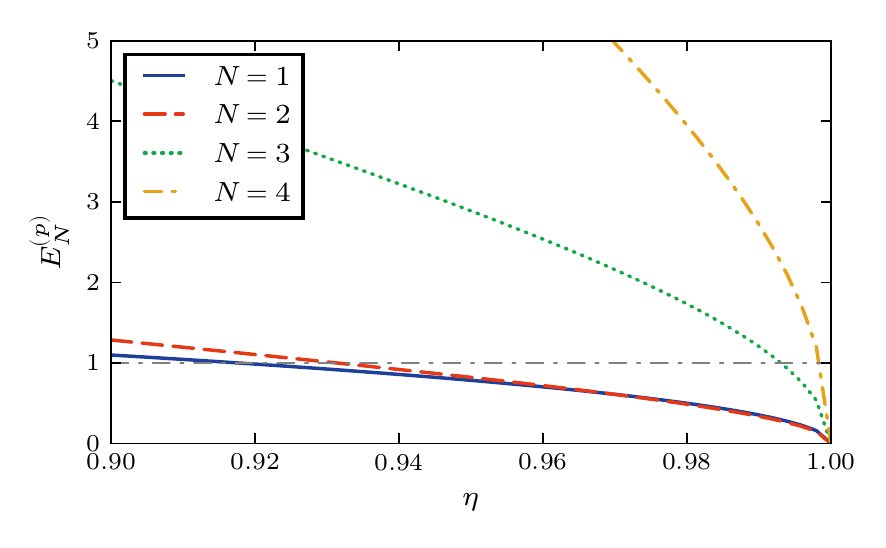}

\caption{\textbf{\emph{Predictions for EPR steering of the NOON states. }}EPR
steering is observed when $E_{N}^{(p)}<1$. It is assumed that the
two-mode NOON state is created and that each mode is then (independently)
subjected to losses. Loss at each mode is modeled by a beam splitter
coupling as described in Sec. VI. Here the beam splitter transmission
efficiencies are $\eta=\eta_{a}=\eta_{b}$\textbf{\textcolor{red}{}}.
\textcolor{green}{}\textcolor{black}{We select the NOON state (\ref{eq:noon-state})
with $\phi=0$ for $N$ odd and $\phi=\pi/2$ for $N$ even. }}
\end{figure}

\textcolor{black}{In Section VI, we will evaluate predictions for
non-ideal case where loss is present.} To complete the prediction
for the steering inequalities with loss present, we also need to calculate
$\Delta_{inf}(P^{N})$, $\Delta_{inf}(X^{N})$. In this paper, we
use $\hat{O}_{x}=\hat{O}_{p}=\hat{X}_{a}$ as the measurement on mode
$a$. As above, we take $x$ to be the result of the measurement $X_{a}$.
We evaluate \textbf{\textcolor{red}{}} 
\begin{eqnarray}
\Delta_{inf}^{2}(P_{b}^{N}) & = & \intop_{-\infty}^{\infty}P\left(x\right)\{\Delta(P_{b}^{N}|x)\}^{2}\,dx\nonumber \\
 & = & \intop_{-\infty}^{\infty}P\left(x\right)[\langle P_{b}^{2N}\rangle_{x}-\langle P_{b}^{N}\rangle_{x}^{2}]\,dx\nonumber \\
\label{eq:PinfN}
\end{eqnarray}
\textcolor{black}{where $\langle...\rangle_{x}$} denotes the expectation
value conditioned on the result $x$, as defined for (\ref{eq:defnzcond}).
The $\langle P_{b}^{2N}\rangle_{x}$ and $\langle P_{b}^{N}\rangle_{x}$
can be expressed in terms of the momentum representation functions
$\langle p|N\rangle$ as shown in Appendix \ref{sec:AppendixInferredVariancesN_NL}.
Similarly 
\begin{eqnarray}
\Delta_{inf}^{2}(X_{b}^{N}) & = & \intop_{-\infty}^{\infty}P\left(x\right)\{\Delta(X_{b}^{N}|x)\}^{2}\,dx\nonumber \\
 & = & \intop_{-\infty}^{\infty}P\left(x\right)\left[\langle X_{b}^{2N}\rangle_{x}-\langle X_{b}^{N}\rangle_{x}^{2}\right]\,dx\nonumber \\
\label{eq:XinfN}
\end{eqnarray}
The \textcolor{black}{$\langle X_{b}^{N}\rangle_{x}$ and }$\langle X_{b}^{2N}\rangle_{x}$
can be solved in terms of the harmonic oscillator wavefunctions $\langle x|N\rangle$
(\ref{eq:herm}) as shown in Appendix \ref{sec:AppendixInferredVariancesN_NL}
and explained for $N=1$, $2$ below.  We have introduced the shorthand
notation $\Delta^{2}x\equiv(\Delta x)^{2}$ to avoid overuse of brackets.
We have solved for the effect of loss on the NOON states using the
methods outlined in Section VI and the results for the steering inequalities
are plotted in Figure 1.

\section{Special case of $N=1$}

Steering for the case of $N=1$ has been proposed by Jones and Wiseman
\cite{JWone photon} and experimentally achieved by Fuwa et al \cite{Furusawa}.
The inequalities used in those papers verified steering in the high
efficiency limit based on homodyne detection, thus giving a firm experimental
proof of the nonlocality of the NOON ($N=1$) state. Here, we outline
the application of the steering inequalities \textcolor{black}{(\ref{eq:steerineqNP-1-1})}
and \textcolor{black}{(\ref{eq:steer_x^N-2})} for this case. 

For $N=1$, the relevant Heisenberg uncertainty relations are\textcolor{red}{{}
}$\Delta n\Delta P\geq|\langle X\rangle|/2$ and $\Delta n\Delta X\geq|\langle P\rangle|/2$\textcolor{black}{{}
. }We see from (\ref{eq:steerthree}) that a criterion sufficient
to certify EPR steering of mode $b$ by measurements on mode $a$
is 
\begin{equation}
\Delta_{inf}n_{b}\Delta_{inf}P_{b}<|\langle X_{b}\rangle|_{inf}/2\label{eq:steeringcritP1}
\end{equation}
The inequality $\Delta_{inf}n_{b}\Delta_{inf}X_{b}<|\langle P_{b}\rangle|_{inf}/2$
is also a steering criterion. Note we can also define the corresponding
criteria for steering of the $a$ mode by interchanging the $a$ and
$b$ indices. The quantities have been defined above in Section II
and III. 

The choice of measurements $\hat{O}_{n}$, $\hat{O}_{c}$, $\hat{O}_{p}$,
$\hat{O}_{x}=\hat{X}_{a}$ to be made on the mode $a$ (as defined
for equations (\ref{eq:steerineqNP-1-1}) and (\ref{eq:steer_x^N-2}))
is generally so as to optimise the criterion for a given state, but
is otherwise not explicitly specified in the criterion. Here, the
choice of $\hat{O}_{n}=\hat{n}_{a}$ is crucial because it takes advantage
of the correlation of number between the two modes of the NOON state,
to allow precisely that $\Delta_{inf}n_{b}=0$. The criterion (\ref{eq:steeringcritP1})
is then predicted to be satisfied for any finite $\Delta_{inf}P_{b}$,
provided $|\langle X_{b}\rangle|_{inf}\neq0$. For the choice of $\hat{O}_{c}$,
$\hat{O}_{p}$, we focus on quadrature phase amplitude measurements
because they are readily measurable experimentally. For $\Delta_{inf}P_{b}$,
we select $\hat{O}_{p}=\hat{X}_{a}$, but we note in the Appendices
(B-D) that the result is not particularly sensitive to this choice.
On the other hand, without a suitable measurement on mode $a$, $|\langle X_{b}\rangle|_{inf}$
will vanish. We find below that the measurement $\hat{X}_{a}$ on
$a$ does not completely collapse the state $b$, and the resulting
superposition predicts a nonzero result for $|\langle X_{b}\rangle|_{inf}$.
With this motivation, we take $\hat{O}_{n}=\hat{n}_{a}$, $\hat{O}_{p}=\hat{X}_{a}$,
$\hat{O}_{x}=\hat{X}_{a}$ and $\hat{O}_{c}=\hat{X}_{a}$. 

We examine the NOON state $|\psi\rangle=\frac{1}{\sqrt{2}}\{|N\rangle|0\rangle+|0\rangle|N\rangle\}$
($\phi=0$) and restrict therefore to the steering criterion (\ref{eq:steeringcritP1}).
The measurement of $n_{a}$ will enable a perfectly accurate prediction
for the number $n_{b}$, so that $\Delta_{inf}n=0$. Taking $\hat{O}_{c}=X_{a}$
we evaluate the mean of $X_{b}$ (or $P_{b}$) at $b$, given a result
$x$ for measurement of $X_{a}$ at $A$. This enables us to evaluate
$|\langle X_{b}\rangle|_{inf}$ and $|\langle P_{b}\rangle|_{inf}$
for a valid steering criterion. If we measure $X_{a}$ with result
$x$, the normalised reduced wave function is (we denote the eigenstate
of $X$ for mode $a$ by $|x\rangle$)
\begin{equation}
\vert\psi\rangle_{x}=\frac{\langle x|N\rangle|0\rangle+\langle x|0\rangle|N\rangle}{\sqrt{|\langle x|N\rangle|^{2}+|\langle x|0\rangle|^{2}}}\label{eq:statered}
\end{equation}
Thus we write the reduced density operator as 
\begin{eqnarray}
\rho_{red,x} & = & \frac{1}{2P(x)}\{|\langle x|N\rangle|^{2}|0\rangle\langle0|+|\langle x|0\rangle|^{2}|N\rangle\langle N|\nonumber \\
 &  & +\langle0|x\rangle\langle x|N\rangle|0\rangle\langle N|+\langle N|x\rangle\langle x|0\rangle|N\rangle\langle0|\}\nonumber \\
\label{eq:rhored}
\end{eqnarray}
where the probability distribution for obtaining a result $x$ for
$X_{a}$ is 
\begin{equation}
P(x)=\frac{1}{2}\{|\langle x|0\rangle|^{2}+|\langle x|N\rangle|^{2}\}\label{eq:Px}
\end{equation}
Here $\langle x|n\rangle$ are the standard oscillator wave functions\textcolor{blue}{{}
}\textcolor{black}{
\begin{equation}
\left\langle x\vert n\right\rangle =\left(\sqrt{\pi}2^{n}n!\right)^{-\frac{1}{2}}\frac{2^{\frac{1}{4}}}{\sqrt{c}}e^{-\frac{x^{2}}{c^{2}}}H_{n}\left(\frac{\sqrt{2}}{c}x\right)\label{eq:herm}
\end{equation}
}involving Hermite polynomials $H_{n}$ and derived \textcolor{black}{using
that $\hat{x}=\frac{c}{2}\left(\hat{a}+\hat{a}^{\dagger}\right),\hat{p}=\frac{c}{2i}\left(\hat{a}-\hat{a}^{\dagger}\right)$}.
In this paper we have taken $c=2$. Now we see that the mean for $X_{b}$
given the result $x$ for $X_{a}$ is 
\begin{eqnarray}
\langle X_{b}\rangle_{x} & = & Tr(\rho_{red,x}X_{b})\nonumber \\
 & = & \frac{1}{2P(x)}\{\langle0|x\rangle\langle x|N\rangle\langle N|X_{b}|0\rangle\nonumber \\
 &  & +\langle N|x\rangle\langle x|0\rangle\langle0|X_{b}|N\rangle\}\label{eq:6}
\end{eqnarray}
and similarly 
\begin{eqnarray}
\langle P_{b}\rangle_{x} & = & Tr(\rho_{red,x}P_{b})\nonumber \\
 & = & \frac{1}{2P(x)}\{\langle0|x\rangle\langle x|N\rangle\langle N|P_{b}|0\rangle|\nonumber \\
 &  & +\langle N|x\rangle\langle x|0\rangle\langle0|P_{b}|N\rangle\}\label{eq:7}
\end{eqnarray}
In fact the mean $\langle X_{b}\rangle_{x}$ will be nonzero only
for\textbf{ $N=1$}, in which case the steering criterion (\ref{eq:steeringcritP1})
is satisfied because $\Delta_{inf}n_{b}=0$ (and $\Delta_{inf}P_{b}\neq\infty$).
Hence, the inequality (\ref{eq:steeringcritP1}) is a suitable steering
criterion for $N=1$. Specifically, following the definition (\ref{eq:defnzcond}),
we evaluate 

\begin{eqnarray}
|\langle X_{b}\rangle|_{inf} & = & \intop_{-\infty}^{\infty}P(x)|\langle X_{b}\rangle_{x}|dx=\sqrt{\frac{2}{\pi}}\label{eq:xint}
\end{eqnarray}
\textcolor{black}{where $\langle X_{b}\rangle_{x}$ is the conditional
quantity between two modes, as defined in (\ref{eq:defnzcond}). }\textcolor{green}{}To
complete the prediction for the steering inequality, we calculate
a suitable value for $\Delta_{inf}P_{b}$ by selecting the measurement
at $A$ to be $X_{a}$. We denote the result of that measurement by
$x$. Then the reduced density operator is $\rho_{red,x}$ as above,
which for $N=1$ gives 

\begin{eqnarray}
(\Delta(P_{b}|x))^{2} & = & \frac{1}{2P\left(x\right)}\{|\langle x|1\rangle|^{2}+3|\langle x|0\rangle|^{2}\}\label{eq:p5}
\end{eqnarray}
and thus
\begin{eqnarray}
\Delta_{inf}^{2}P_{b} & = & \intop_{-\infty}^{\infty}P(x)\{\Delta(P_{b}|x)\}^{2}dx\nonumber \\
 & = & \frac{1}{2}\intop_{-\infty}^{\infty}\{|\langle x|1\rangle|^{2}+3|\langle x|0\rangle|^{2}\}=2\label{eq:p8}
\end{eqnarray}
where $\Delta^{2}x\equiv(\Delta x)^{2}$. We obtain an EPR steering
when $E_{1}^{(p)}\equiv\frac{\Delta_{inf}n_{b}\Delta_{inf}P_{b}}{|\langle X_{b}\rangle|_{inf}/2}<1$.
For the ideal NOON state with no losses, $E_{1}^{(p)}=0$ and the
steering is always detectable via this criterion. The situation with
loss is studied in Section VI and presented in Figure 1. Efficiencies
$\eta>0.92$ are required to detect the steering.

\section{Special case of $N=2$}

\emph{}We now examine the details for the NOON state with $N=2$
which represents an important case potentially accessible to experiment,
in view of recent advances \cite{atom hongmandel,krusesilberhorneexp,Furusawa}.
Firstly, $[n,X^{2}]=-i(XP+PX)=-2iXP-2=2(a^{\dagger2}-a^{2}).$ Similarly,
$[n,P^{2}]=i(XP+PX)=-2(a^{\dagger2}-a^{2})$.  \textcolor{black}{The
steering criteria are}

\textcolor{black}{
\begin{eqnarray}
\Delta_{inf}n\Delta_{inf}(X^{2}) & < & |\langle a^{\dagger2}-a^{2}\rangle|_{inf}\label{eq:steerP3-1}
\end{eqnarray}
and }\textcolor{green}{}\textcolor{black}{{} }

\textcolor{black}{
\begin{eqnarray}
\Delta_{inf}n\Delta_{inf}(P^{2}) & < & |\langle a^{\dagger2}-a^{2}\rangle|_{inf}\label{eq:steerP3-1-1}
\end{eqnarray}
 }For the NOON state (\ref{eq:noon-state}) with $N=2$ we obtain
\begin{eqnarray*}
|\langle a^{\dagger2}-a^{2}\rangle|_{inf} & = & \intop P\left(x\right)|\langle\hat{a}^{2}|x\rangle-\langle\left(\hat{a}^{\dagger}\right)^{2}|x\rangle|\,dx\\
 & = & \sqrt{2}|\sin\phi|\intop|\langle x|0\rangle\langle x|2\rangle|\,dx\\
 & = & 2\sqrt{\frac{2}{e\pi}}|\sin\phi|=0.968|\sin\phi|
\end{eqnarray*}
\textcolor{red}{}Both the steering criteria (\ref{eq:steerP3-1})
and (\ref{eq:steerP3-1-1}) become useful for the NOON state with
$\phi=\pi/2$. We show in Appendix B by integration of the Hermite
polynomials that $\Delta_{inf}(X^{2})=3.18$ and $\Delta_{inf}(P^{2})=3.18$.\textcolor{green}{}
For the ideal case with no detection loss, $\Delta_{inf}n=0$ and
the steering for the NOON state with $N=2$ is detectable using either
criterion. Comparing with the results for $N=1$, we see that the
prediction for the ratio of the right- to left-sides of the steering
inequalities decreases for $N=2$. We expect the criteria will be
more difficult to satisfy at higher $N$ in non-ideal cases. Details
of the calculations for arbitrary $N$ are given in Appendices B and
C. The effect of the losses is studied below in Section VI and the
results are shown in Figure 1.\textcolor{blue}{}

\section{Including losses}

Signatures of the NOON state superposition are known to be fragile
to losses. We examine the effect of loss on the signatures proposed
here, by using a simple model for loss. We couple each mode $a$ and
$b$ to second independent fields taken as single modes and initially
in independent vacuum states, following the beam splitter model introduced
for the study of the decoherence of a macroscopic superposition state
by Yurke and Stoler \cite{ystbeamsplit-1}. We thus evaluate the moments
of detected fields with boson operators $a_{det}$, $b_{det}$ given
by\textcolor{black}{{} 
\begin{eqnarray}
a_{det} & = & \sqrt{\eta_{a}}a+\sqrt{1-\eta_{a}}a_{v}\nonumber \\
b_{det} & = & \sqrt{\eta_{b}}b+\sqrt{1-\eta_{b}}b_{v}\label{eq:loss-1}
\end{eqnarray}
}Here the $a_{v}$ and $b_{v}$ are destruction operators for independent
external vacuum modes that couple to the modes of the NOON state.
These external modes model the presence of an external environment
into which quanta can be lost from the $a$ and $b$ modes. The amount
of coupling for each mode is determined by the efficiency factors
$\eta_{a}$ and $\eta_{b}$ respectively. The $\eta_{A/B}=1$ indicates
zero loss; low $\eta_{A/B}$ indicates high loss. The model is effective
for optical NOON states where thermal noise can be neglected. The
full calculation is explained in Appendix \ref{sec:SteeringCriteria_Losses}.
We find for $N=1$ \textbf{\textcolor{red}{}}\textcolor{black}{and
$\phi=0$ }

\begin{eqnarray}
E_{1}^{(p)} & \equiv & \frac{\Delta_{inf}n\Delta_{inf}P}{|\langle X\rangle|_{inf}/2}\nonumber \\
 & = & 2\left[\frac{\eta_{b}\left(\eta_{a}+\eta_{b}-2\right)}{2\left(\eta_{a}-2\right)}\left(1+\eta_{b}\right)\right]^{\frac{1}{2}}/\left[\sqrt{\frac{2}{\pi}}\sqrt{\eta_{a}\eta_{b}}\right]\nonumber \\
\label{eq:E1p_losses_phi0}
\end{eqnarray}
\textcolor{red}{}\textcolor{green}{}The expressions for higher
$N$ are more complex but are explained in Appendix \ref{sec:SteeringCriteria_Losses}
and evaluated numerically.\textcolor{red}{{} }\textcolor{blue}{}Figure
1 shows $E_{N}^{(p)}$ versus $\eta$, for the case of symmetrical
efficiency $\eta=\eta_{a}=\eta_{b}$. The criterion for EPR steering
is satisfied for $N=1$ provided $\eta>0.92$ but as expected for
the NOON state, the cut-off efficiency increases sharply for higher
$N$.\textcolor{red}{{} }For $N=2$ there is asymmetrical dependence
on $\eta_{a}$ and $\eta_{b}$ as evident by the contour plots of
Figure 2. The signature appears more sensitive to the efficiency $\eta_{B}$
of mode $b$. Such asymmetrical sensitivity depending on the steering
direction has been noted previously \cite{asymsteer,thermal qm}.

\begin{figure}

\includegraphics{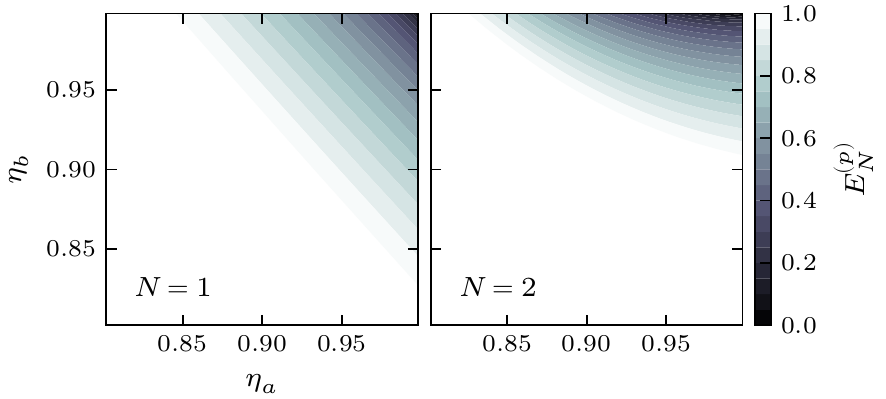}

\caption{\textbf{\emph{Contour plot shows the effect of loss on the EPR steering:
}}EPR steering is observed when $E_{N}^{(p)}<1$. The $\eta_{a}$
and $\eta_{b}$ are the efficiencies for detection of mode $a$ and
$b$ respectively\textcolor{black}{.} \textcolor{green}{ }}
\end{figure}

We note that the model (\ref{eq:loss-1}) describes losses that occur
\emph{prior }to detection. It is assumed that the subsequent detection
process gives no further loss. Alternatively, if the beam splitter
is to model detection losses, then the losses would need to be assumed
\emph{identical }for each of the detection processes (number or homodyne).
In reality, for low $N$ the numbers $n_{a}$, $n_{b}$ are usually
detected via counting techniques where the efficiency of detection
is often small. On the other hand, the quadratures $X$ and $P$ are
measured via homodyne detection where efficiencies are high (at least
for optical fields). This creates a situation where the loss coefficient
$\eta$ is dependent on the choice of measurement made at each site,
which we point out can \emph{create loopholes} in the use of the signature
for a practical experiment if not considered carefully \cite{Bell-1,loopholdet}.\textcolor{green}{{}
}\textcolor{black}{We discuss this further in the Conclusion.}

\section{Measurement}

We next consider how to experimentally \emph{measure} the moments
on the right side of the steering inequalities \textcolor{black}{(\ref{eq:steerineqNP-1-1})}
and \textcolor{black}{(\ref{eq:steer_x^N-2})}. For $N=1$ this is
straightforward as explained in Section IV. For $N=2$, on examining
the expressions (\ref{eq:eqncal-1}) and (\ref{eq:eqncal2}), we see
we need to measure\textcolor{black}{{} $\langle\left[n,P^{2}\right]\rangle=\langle XP+PX\rangle$.
}We define the measurable rotated quadrature phase amplitudes as $X_{\theta}=X\cos(\theta)+P\sin(\theta)$
and $P_{\theta}=-X\sin(\theta)+P\cos(\theta)$. Hence, $X_{\pi/4}=\frac{1}{\sqrt{2}}\{X+P\}$
and $P_{\pi/4}=\frac{1}{\sqrt{2}}\{-X+P\}$ and we note that $\langle X_{\pi/4}^{2}\rangle=\langle X^{2}+P^{2}+XP+PX\rangle/2$.
Thus, we can deduce either $\langle XP\rangle$ or $\langle PX\rangle$
by measuring the moments $\langle X^{2}\rangle$, $\langle P^{2}\rangle$
and $\langle X_{\pi/4}^{2}\rangle$. The steering criteria (\ref{eq:steerineq1p-1}-\ref{eq:steerineq2x-1})
for $N=2$ can be written as (here we drop the subscripts $b$ for
convenience)\textcolor{black}{{} }\textcolor{green}{}\textcolor{black}{{} }

\textcolor{black}{
\begin{eqnarray}
\Delta_{inf}n\Delta_{inf}(P^{2}) & < & |\langle\left[n,P^{2}\right]\rangle|_{inf}/2\nonumber \\
 & = & |\langle X_{\pi/4}^{2}-X^{2}/2-P^{2}/2\rangle|_{inf}\nonumber \\
\label{eq:steerpn=00003D2}
\end{eqnarray}
}and

\textcolor{black}{
\begin{eqnarray}
\Delta_{inf}n\Delta_{inf}(X^{2}) & < & |\langle\left[n,X^{2}\right]\rangle|_{inf}/2\nonumber \\
 & = & |\langle X_{\pi/4}^{2}-X^{2}/2-P^{2}/2\rangle|_{inf}\nonumber \\
\label{eq:steerXn=00003D2}
\end{eqnarray}
}The moments of $X$, $P$ and $X_{\pi/4}$ are each measurable using
homodyne detection. 

For $N=3$, we see from (\ref{eq:eqncal-1}) and (\ref{eq:eqncal2})
that we need to measure $\left[n,P^{3}\right]=3\langle P^{2}X+2iP\rangle$
the other measurements being straightforward. Expanding gives\textcolor{red}{}
\textcolor{green}{}\textcolor{green}{}
\begin{eqnarray*}
\langle X_{\pi/4}^{3}-P_{\pi/4}^{3}\rangle & = & \frac{1}{\sqrt{2}}(\langle X^{3}\rangle+6i\langle P\rangle+3\langle P^{2}X\rangle)
\end{eqnarray*}
Hence we can measure $\langle X_{\pi/4}^{3}\rangle,\langle P_{\pi/4}^{3}\rangle$,
$\langle X^{3}\rangle$, $\langle P\rangle$ and consequently infer
the value of $\langle P^{2}X\rangle$. Specifically, the steering
inequalities become

\textcolor{black}{
\begin{eqnarray}
\Delta_{inf}n\Delta_{inf}(P^{3}) & < & |\langle\left[n,P^{3}\right]\rangle|_{inf}/2\nonumber \\
 & = & |\langle\sqrt{2}\left(X_{\pi/4}^{3}-P_{\pi/4}^{3}\right)-X^{3}\rangle|_{inf}/2\nonumber \\
\label{eq:steerpn=00003D3}
\end{eqnarray}
}and 

\textcolor{black}{
\begin{eqnarray}
\Delta_{inf}n\Delta_{inf}(X^{3}) & < & |\langle\left[n,X^{3}\right]\rangle|_{inf}/2\nonumber \\
 & = & |\langle\sqrt{2}\left(X_{\pi/4}^{3}+P_{\pi/4}^{3}\right)-P^{3}\rangle|_{inf}/2\nonumber \\
\label{eq:steerxn=00003D3}
\end{eqnarray}
}We comment that the inequalities (\ref{eq:steerpn=00003D2}-\ref{eq:steerxn=00003D3})
are valid as a sufficiency test of EPR steering for \emph{all }states
i.e. we do not assume ideal NOON states. 

\section{Discussion }

To conclude, we discuss an obvious question, which is how to ensure
in an experiment that the observed steering is due to the quantum
coherence of the NOON superposition, as opposed to an alternative
microscopic effect that might arise from superpositions of number
states distinct by less than $N$ quanta? This is an important question
where losses are present, because then the outcomes for number measurements
can be different to $0$ and $N$.

First, the answer is clear in the ideal case of a two-mode system
that generates only outcomes $0$ or $N$ for the number measurements.
The density operator can then be written in terms of four basis states
$|0\rangle|0\rangle,|0\rangle|N\rangle,|N\rangle|0\rangle,|N\rangle|N\rangle$.
The violation of an EPR steering inequality is also confirmation of
an entangled state, and in this case that can only imply entanglement
involving the mesoscopically distinguishable basis states. The violation
of the steering inequality confirms the presence of an $N$th order
off-diagonal matrix element (i.e. $\langle0|\langle N|\rho|0\rangle|N\rangle\neq0$).
The details are straightforward and given in the Appendix E.

In experiments where loss or noise is present, the distribution $p_{n}$
for number $\hat{n}_{b}$ will include outcomes other than $0$ and
$N$. It is not then clear whether an observation of EPR steering
is a result of the superposition of states such as $|M'\rangle|n\rangle$
and $|M\rangle|m\rangle$ where $M-M'\sim N$, or the result of less
interesting superpositions where $M\sim M'$. 

The problem of determining whether the system has an $N$th order
quantum coherence (defined as $\langle0|\langle N|\rho|0\rangle|N\rangle\neq0$)
is nontrivial \cite{ericpramix,cavalreidgerd,macro-coh_verdral,frowis_3}.
However the following approach based on the steering inequality may
be useful. The outcomes for number at mode $a$ are either $n_{a}>0$
or $n_{a}=0$. The distribution for the outcome $n_{b}$ of number
at mode $b$ given any result $n_{a}>0$ is (for small losses) a ``hill''
\foreignlanguage{australian}{centred} near (or at) $0$. The distribution
for $n_{b}$ given the result $n_{a}=0$ is a ``hill \foreignlanguage{australian}{centred}
near $N$. The mean and variance of each of the two hills is measurable
and denoted by $\langle n_{b}\rangle_{1}$, $(\Delta\hat{n}_{b})_{1}^{2}$
and $\langle n_{b}\rangle_{2}$, $(\Delta\hat{n}_{b})_{2}^{2}$.
For small losses, each of the two variances will be \emph{small}.

We suppose that the experimentalist has measured a violation of the
EPR steering inequality $\Delta_{inf}\hat{n}_{b}\Delta_{inf}(\hat{P}_{b}^{N})\geq|\langle\hat{C}_{b}\rangle|_{inf}/2$
where $(\Delta_{inf}\hat{n}_{b})^{2}=\sum_{n_{a}}P(n_{a})(\Delta(n_{b}|n_{a}))^{2}$
and $P(n_{a})$ is the probability of outcome $n_{a}$. We note that
where the conditional distribution for $n_{b}$ given $n_{a}$ is
uniform for all $n_{a}>0$, this will imply violation of the new but
similar inequality 
\begin{equation}
\Bigl(P_{1}(\Delta\hat{n}_{b})_{1}^{2}+P_{2}(\Delta\hat{n}_{b})_{2}^{2}\Bigr)(\Delta_{inf}\hat{P}_{b}^{N})^{2}\geq\frac{1}{4}|\langle\hat{C_{b}}\rangle|_{inf}^{2}\label{eq:uncertwomode-1}
\end{equation}
Here we specify as selected in (\ref{eq:steerinequintro}) that the
inferred values for $\hat{P_{b}}^{N}$and $\hat{C}_{b}$ are calculated
using the same observable at mode $a$. Here $P_{1}$ is the probability
of $n_{a}>0$ and we have assumed $(\Delta\hat{n}_{b})_{1}^{2}=\Delta(n_{b}|n_{a})$
for $n_{a}>0$ . Similarly, $P_{2}$ is the probability that $n_{a}=0$
and $(\Delta\hat{n}_{b})_{2}^{2}=\Delta(n_{b}|n_{a})$ for $n_{a}=0$.
We note that the loss model of Section VI predicts the distributions
to be uniform, but if this is not the case then the inequality can
be measured directly. 

It is shown in the Appendix F that violation of the inequality (\ref{eq:uncertwomode-1})
is a negation of the mixture 
\begin{equation}
\rho=P_{1}\rho_{1}^{ab}+P_{2}\rho_{2}^{ab}\label{eq:twomodemix}
\end{equation}
where $\rho_{1}^{ab}$ and $\rho_{2}^{ab}$ are two-mode density operators
with a mean and variance for $\hat{n}_{b}$ given by $\langle n_{b}\rangle_{1}$,
$(\Delta n_{b})_{1}^{2}$ and $\langle n_{b}\rangle_{2}$, $(\Delta n_{b})_{2}^{2}$
respectively. The negation is for\emph{ all} mixtures of the form
(\ref{eq:twomodemix}), which includes where $\rho_{i}^{ab}$ can
be a superposition of number states. However the spread of number
states involved in the superposition is constrained by the small variances
associated with each $\rho_{i}^{ab}$. The $\rho_{1}^{ab}$ and $\rho_{2}^{ab}$
each have a variance for $\hat{n}_{b}$ that is narrower than the
variance of the distribution given by the NOON superposition state.
In other words, the violation of the inequality (\ref{eq:uncertwomode-1})
can only be consistent with a density operator $\rho$ involving superpositions
$|\psi_{sup}\rangle$ of states distributed over \emph{both }hills.

\section{Conclusion}

The particular steering inequalities we present in this paper involve
measurements of number as well as quadrature phase amplitude correlation.
Number measurements often entail poor efficiencies. It would seem
feasible to perform in the first instance an experiment based on post-selection
of the events where a total of $N$ quanta (e.g. photons) are detected
across both sites. The problem of distinguishing multiple from single
photon counts at a given location require photon number-resolving
detectors, or could be handled with $N$-photon counts being evaluated
using multiple beam splitters \cite{krusesilberhorneexp,NOON_N5}.

The experiment for $N=2$ would be a demonstration of a higher order
(more mesoscopic) nonlocality than for $N=1$ and would seem not unrealistic
given the high efficiencies available with homodyne detection. Our
calculations show that $\eta>0.94$ is required. Care is needed to
model the homodyne inefficiency as a \emph{loss before detection},
and this small amount of loss must therefore \emph{also} enter into
the evaluation of the number correlation, to avoid the well-documented
possible loopholes associated with losses that depend on measurement
choices. The experiment for $N=1$ is feasible. Such an experiment
would complement that performed recently by Fuwa et al \cite{Furusawa}
based on a different EPR steering inequality. 

Finally, we point out that the steering inequalities (\ref{eq:steerineq1p-1}-\ref{eq:steerineq2x-1})
might be useful for detecting steering in other two-mode systems,
especially where there is an inter-mode photon number correlation
so that $\Delta_{inf}n_{b}=0$. For instance, we can apply the first
order inequality $\Delta_{inf}\hat{n}_{b}\Delta_{inf}\hat{P}_{b}<|\langle\hat{X}_{b}\rangle|_{inf}/2$
(Eq. (\ref{eq:steeringcritP1-1})) to the two-mode squeezed state.
Denoting the two-mode squeeze parameter by $r$, the solutions for
this state give $\Delta_{inf}n_{b}=0$ for all $r$. Further, it is
well known that there is an EPR correlation between the quadrature
phase amplitudes of the two modes for all $r$ \cite{eprr-2,rrmp},
so that $|\langle\hat{X}_{b}\rangle|_{inf}\neq0$ and $\Delta_{inf}\hat{P}_{b}\rightarrow0$
as $r\rightarrow\infty$. While steering has been experimentally achieved
for this state via the alternative EPR steering inequality $\Delta_{inf}\hat{X}_{b}\Delta_{inf}\hat{P}_{b}<1$
\cite{cvepr_ou-1,eprr-2,eprrdrum}, it is quite possible that the
use of the steering inequality with the number correlation $\Delta_{inf}n_{b}=0$
(which is valid for all $r$) may provide advantages in some regimes.
\begin{acknowledgments}
We thank the Australian Research Council for support through the Discovery
Project Grants scheme. We also thank Bryan Dalton, Peter Drummond
and M. Chekhova for stimulating discussions about NOON states.
\end{acknowledgments}

\appendix

\section{Proof of Result (1)\label{sec:AppendixProofs}}

We will assume that the LHS model holds, for which moments are given
by

\begin{eqnarray}
\langle X_{A}(\theta)X_{B}(\phi)\rangle & = & \int_{\lambda}d\lambda P(\lambda)\langle X_{A}(\theta)\rangle_{\lambda,\rho}\langle X_{B}(\phi)\rangle_{\lambda}\nonumber \\
 & \equiv & \sum_{R}P_{R}\langle X_{A}(\theta)\rangle_{R,\rho}\langle X_{B}(\phi)\rangle_{R}\label{eq:lhs-1}
\end{eqnarray}
Here we give two alternative (but equivalent) notations for the hidden
variable-type parameters, denoting the continuous variable option
by the symbol $\lambda$ as in Bell's work and the discrete option
by $R$. The proof is unchanged whether we use integrals ($\lambda)$
or discrete summations ($R$). 

We consider the inference variance $(\Delta_{inf}\sigma_{A}^{X})^{2}$.
Based on the definitions given in Section III, we see that $\sum_{x_{j}^{B}}P(x_{j}^{B})\{\Delta(\sigma_{A}^{X}|x_{j}^{B})\}^{2}=\sum_{x_{j}^{B}}P(x_{j}^{B})\sum_{\sigma_{A}^{X}}P(\sigma_{A}^{X}|x_{j}^{B})\{\sigma_{A}^{X}-\langle\sigma_{A}^{X}\rangle_{x_{j}^{B}}\}^{2}$
which we can re-express as $\sum_{x_{j}^{B},\sigma_{A}^{X}}P(x_{j}^{B},\sigma_{A}^{X})\{\sigma_{A}^{X}-\langle\sigma_{A}^{X}\rangle_{x_{j}^{B}}\}^{2}$
and hence as $\sum_{R}P_{R}\sum_{x_{j}^{B},\sigma_{A}^{X}}P_{R}(x_{j}^{B},\sigma_{A}^{X})\{\sigma_{A}^{X}-\langle\sigma_{A}^{X}\rangle_{x_{j}^{B}}\}^{2}$.
This follows using that for a probabilistic (hidden variable) mixture
$P(x_{j}^{B},\sigma_{X}^{A})=\sum_{R}P_{R}P_{R}(x_{j}^{B},\sigma_{X}^{A})$.
Now we note that $\langle(x-\delta)^{2}\rangle\geq\langle(x-\langle x\rangle)^{2}\rangle$
where $\delta$ is any number. Hence the expression becomes bounded
from below, and we can simplify further to show that\textcolor{blue}{{}
}

\begin{eqnarray*}
 &  & \sum_{R}P_{R}\sum_{x_{j}^{B},\sigma_{A}^{X}}P_{R}(x_{j}^{B},\sigma_{A}^{X})\{\sigma_{A}^{X}-\langle\sigma_{A}^{X}\rangle_{x_{j}^{B}}\}^{2}\\
 &  & \geq\sum_{R}P_{R}\sum_{x_{j}^{B}}P_{R}(x_{j}^{B})\{\Delta_{R}(\sigma_{A}^{X}|x_{j}^{B})\}^{2}\\
 &  & =\sum_{R}P_{R}\{\Delta_{inf,R}\sigma_{A}^{X}\}^{2}
\end{eqnarray*}
Here, the subscripts $R$ imply that the probabilities, averages and
variances are with respect to the state $R$ and we have used that
$\{\Delta_{R}(\sigma_{A}^{X}|x_{j}^{B})\}^{2}=\sum_{\sigma_{A}^{X}}P_{R}(\sigma_{A}^{X}|x_{j}^{B})\{\sigma_{A}^{X}-\langle\sigma_{A}^{X}\rangle_{x_{j}^{B},R}\}^{2}$.
We note that the symbol $\lambda$ is used alternatively to $R$ in
the main text, to describe that the variables may also be continuous.
The proof follows similarly in either case. Now, if we assume the
separability between the bipartition $A-B$ for each state $R$, in
accordance with the LHS model (\ref{eq:lhs}), then 
\begin{equation}
P_{R}(x_{j}^{B},\sigma_{A}^{X})=P_{R}(x_{j}^{B})P_{R}(\sigma_{A}^{X})\label{eq:sep-1}
\end{equation}
This implies $\langle\sigma_{A}^{X}\rangle_{x_{j}^{B},R}=\langle\sigma_{A}^{X}\rangle_{R}$
and $\{\Delta_{R}(\sigma_{A}^{X}|x_{j}^{B})\}^{2}=(\Delta_{R}\sigma_{A}^{X})^{2}$.
Then we find, on using $\sum_{x_{j}^{B}}P_{R}(x_{j}^{B})=1$, that
we can write $\{\Delta_{inf,R}\sigma_{A}^{X}\}^{2}=\{\Delta_{R}\sigma_{A}^{X}\}^{2}$.
Thus, on applying the Cauchy-Schwarz inequality, we see that 
\begin{eqnarray*}
\Delta_{inf}^{2}\sigma_{A}^{X}\Delta_{inf}^{2}\sigma_{A}^{Y} & \geq & (\sum_{R}P_{R}\{\Delta_{R}\sigma_{A}^{X}\}^{2})(\sum_{R}P_{R}\{\Delta_{R}\sigma_{A}^{Y}\}^{2})\\
 & \geq & (\sum_{R}P_{R}\{\Delta_{R}\sigma_{A}^{X}\}\{\Delta_{R}\sigma_{A}^{Y}\})^{2}
\end{eqnarray*}
where we define $\Delta_{inf}^{2}\sigma_{A}^{X}\equiv(\Delta_{inf}\sigma_{A}^{X})^{2}$
and 
\[
\Delta_{inf}^{2}\sigma_{A}^{Y}\equiv(\Delta_{inf}\sigma_{A}^{Y})^{2}=\sum_{y_{j}^{B}}P(y_{j}^{B})\{\Delta(\sigma_{A}^{X}|y_{j}^{B})\}^{2}
\]
noting that the $\left\{ y_{j}\right\} $ is the set of results for
a measurement $y$ made at $B$ to infer the value of the measurement
of $\sigma_{A}^{Y}$ at $A$. We consider an LHS model (\ref{eq:lhs})
where we assume the states at $A$ are local \emph{quantum} states,
so that we can use quantum uncertainty relations to derive a final
steering inequality: e.g. $\{\Delta_{R}(\sigma_{A}^{X})\}\{\Delta_{R}(\sigma_{A}^{Y})\}\geq|\langle\sigma_{A}^{Z}\rangle_{R}|/2$
for any quantum state denoted by $R$. Using the above results, the
LHS model implies
\begin{eqnarray*}
(\Delta_{inf}\sigma_{A}^{X})(\Delta_{inf}\sigma_{A}^{Y}) & \geq & \sum_{R}P_{R}\{\Delta_{R}\sigma_{A}^{X}\}\{\Delta_{R}\sigma_{A}^{Y}\}\\
 & \geq & \sum_{R}P_{R}(|\langle\sigma_{A}^{Z}\rangle_{R}|/2)
\end{eqnarray*}
However, for a separable model, we know that $\langle\sigma_{A}^{Z}\rangle_{z_{j}^{B},R}=\langle\sigma_{A}^{Z}\rangle_{R}$
and hence 
\begin{eqnarray*}
\sum_{z_{j}^{B}}P(z_{j}^{B})\sum_{R}P_{R}|\langle\sigma_{A}^{Z}\rangle_{z_{j}^{B},R}| & = & \sum_{R}P_{R}\sum_{z_{j}^{B}}P(z_{j}^{B})|\langle\sigma_{A}^{Z}\rangle_{R}|\\
 & = & \sum_{R}P_{R}|\langle\sigma_{A}^{Z}\rangle_{R}|
\end{eqnarray*}
where here the $\left\{ z_{j}\right\} $ is the set of results for
a measurement $z$ at $B$, that we use to infer results for $\sigma_{A}^{Z}$.
Hence 
\begin{eqnarray*}
(\Delta_{inf}\sigma_{A}^{X})(\Delta_{inf}\sigma_{A}^{Y}) & \geq & \sum_{z_{j}^{B}}P(z_{j}^{B})\sum_{R}P_{R}|\langle\sigma_{A}^{Z}\rangle_{z_{j}^{B},R}|/2\\
 & = & \sum_{z_{j}^{B}}P(z_{j}^{B})|\langle\sigma_{A}^{Z}\rangle_{z_{j}^{B}}|/2
\end{eqnarray*}
 We have used (for states constrained by the LHS model), 
\begin{eqnarray*}
\langle\sigma_{A}^{Z}\rangle_{z_{j}^{B}} & = & \sum_{\sigma_{A}^{Z}}\sigma_{A}^{Z}P(\sigma_{A}^{Z}|z_{j}^{B})\\
 & = & \sum_{\sigma_{A}^{Z}}\sigma_{A}^{Z}\sum_{R}P_{R}P_{R}(\sigma_{A}^{Z}|z_{j}^{B})\\
 & = & \sum_{R}P_{R}\langle\sigma_{A}^{Z}\rangle_{z_{j}^{B},R}
\end{eqnarray*}
Defining $|\langle\sigma_{A}^{Z}\rangle|_{inf}=\sum_{z_{j}^{B}}P(z_{j}^{B})|\langle\sigma_{A}^{Z}\rangle_{z_{j}^{B}}|$,
we see finally that the LHS model implies $(\Delta_{inf}\sigma_{A}^{X})(\Delta_{inf}\sigma_{A}^{Y})\geq|\langle\sigma_{A}^{Z}\rangle|_{inf}/2$.
Violation of this inequality implies failure of the LHS model, and
therefore implies steering of $A$ by $B$. The result is steering
of $B$ by $A$ if the $A$ and \textbf{$B$ }indices are exchanged
(as in the main text). This completes the proof.$\boxempty$

\section{Evaluation of inferred variances \label{sec:AppendixInferredVariancesN_NL}}

Here we will evaluate the inferred uncertainties $\Delta_{inf}(X^{N})$
and $\Delta_{inf}(P^{N})$ for the NOON state given in Eq. (\ref{eq:noon-state}).
We first consider $X\equiv X_{b}$ and evaluate $\Delta_{inf}^{2}(X_{b}^{N})\equiv\left(\Delta_{inf}(X_{b}^{N})\right)^{2}$,
which is given by (\ref{eq:XinfN}). The terms of the form $\left\langle X_{b}^{n}\right\rangle _{inf,x}\equiv\langle X_{b}^{n}|x\rangle$,
with $n=N$ or $n=2N$, are evaluated using the reduced density operator
$\rho_{red,x}$:
\begin{eqnarray}
\rho_{red,x} & = & \frac{1}{2}\{|\langle x|N\rangle|^{2}|0\rangle\langle0|+e^{-i\phi}\langle x|N\rangle\langle0|x\rangle|0\rangle\langle N|\nonumber \\
 &  & +|\langle x|0\rangle|^{2}|N\rangle\langle N|+e^{i\phi}\langle x|0\rangle\langle N|x\rangle|N\rangle\langle0|\}\label{eq:redx}
\end{eqnarray}
 and the fact that operators $\hat{X}$ and $\hat{P}$ can be described
in terms of a complete set of projectors as $\hat{X}_{B}^{n}=\intop_{-\infty}^{\infty}x_{B}^{n}|x_{B}\rangle\langle x_{B}|\,dx_{B}$
and $\hat{P}_{B}^{n}=\intop_{-\infty}^{\infty}p_{B}^{n}|p_{B}\rangle\langle p_{B}|\,dp_{B}$.
Therefore we get:
\begin{eqnarray}
\langle X^{n}\rangle_{inf,x} & = & Tr(\rho_{red,x}X^{n})\nonumber \\
 & = & \frac{1}{2P\left(x\right)}\left[\left|\langle x|N\rangle\right|^{2}\intop x_{B}^{n}\left|\langle x_{B}|0\rangle\right|^{2}dx_{B}\right.\nonumber \\
 &  & +e^{-i\phi}\,\langle x|N\rangle\langle0|x\rangle\intop x_{B}^{n}\langle N|x_{B}\rangle\langle x_{B}|0\rangle\,dx_{B}\nonumber \\
 &  & +e^{i\phi}\,\langle x|0\rangle\langle N|x\rangle\intop x_{B}^{n}\langle0|x_{B}\rangle\langle x_{B}|N\rangle\,dx_{B}\nonumber \\
 &  & \left.+\left|\langle x|0\rangle\right|^{2}\intop x_{B}^{n}\left|\langle x_{B}|N\rangle\right|^{2}\,dx_{B}\right]\label{eq:Xn_N2}
\end{eqnarray}
where $P\left(x\right)=\frac{1}{2}\left[|\langle x|0\rangle|^{2}+|\langle x|N\rangle|^{2}\right]$
is the probability of measuring $X_{A}$ and getting outcome $x$
and $\langle x|N\rangle$ are the harmonic oscillator functions given
in Eq. (\ref{eq:herm}). The value for $\Delta_{inf}(X_{b}^{N})$
is obtained on evaluating the expressions of $\langle X^{n}\rangle_{inf,x}$,
with $n=N$ or $2N$, and substituting on the expression given in
Eq. (\ref{eq:XinfN}). Similarly we evaluate the inferred variance
of $P\equiv P_{b}$, which is given by (\ref{eq:PinfN}). Using the
reduced density operator $\rho_{red,x}$ given above we find:
\begin{eqnarray}
\langle P^{n}\rangle_{inf,x} & = & \frac{1}{2P\left(x\right)}\left[\left|\langle x|N\rangle\right|^{2}\intop p_{B}^{n}\left|\langle p_{B}|0\rangle\right|^{2}\,dp_{B}\right.\nonumber \\
 &  & +e^{-i\phi}\,\langle x|N\rangle\langle0|x\rangle\intop p_{B}^{n}\langle0|p_{B}\rangle\langle p_{B}|N\rangle\,dp_{B}\nonumber \\
 &  & +e^{i\phi}\,\langle x|0\rangle\langle N|x\rangle\intop p_{B}^{n}\langle N|p_{B}\rangle\langle p_{B}|0\rangle\,dp_{B}\nonumber \\
 &  & \left.+\left|\langle x|0\rangle\right|^{2}\intop p_{B}^{n}\left|\langle p_{B}|N\rangle\right|^{2}\,dp_{B}\right]\label{eq:PnNL}
\end{eqnarray}

To evaluate, we first consider $N=2$. We let $\phi=\pi/2$: 
\begin{eqnarray*}
\langle X^{2}\rangle_{inf,x} & = & \frac{1}{2P\left(x\right)}\left[\left|\langle x|2\rangle\right|^{2}\intop x_{B}^{2}\left|\langle x_{B}|0\rangle\right|^{2}dx_{B}\right.\\
 &  & \left.+\left|\langle x|0\rangle\right|^{2}\intop x_{B}^{2}\left|\langle x_{B}|2\rangle\right|^{2}\,dx_{B}\right]\\
 & = & 1+\frac{8}{3-2x^{2}+x^{4}}
\end{eqnarray*}
and 
\begin{eqnarray*}
\langle X^{4}\rangle_{inf,x} & = & 3+\frac{72}{x^{4}-2x^{2}+3}
\end{eqnarray*}
where we have used that $P(x)=\frac{e^{-\frac{x^{2}}{2}}}{2\sqrt{2\pi}}\left(\frac{\left(2x^{2}-2\right)^{2}}{8}+1\right)$.
On performing the integration using the above results we get for the
$N=2$ state that  \textcolor{black}{$\Delta_{inf}^{2}(X_{b}^{2})=10.1351$
and $\Delta_{inf}(X_{b}^{2})=3.18356$.} \textcolor{red}{}Similarly
we evaluate $\Delta_{inf}^{2}(P_{b}^{N})$:
\begin{eqnarray*}
\langle P^{2}\rangle_{inf,x} & = & 1+\frac{8}{3-2x^{2}+x^{4}}\\
\langle P^{4}\rangle_{inf,x} & = & 3+\frac{72}{x^{4}-2x^{2}+3}
\end{eqnarray*}
These results are the same as for $X$, since for this value of angle
$e^{i\phi}=i=-e^{-i\phi}$, and also 
\begin{eqnarray*}
\intop p_{B}^{n}\langle0|p_{B}\rangle\langle p_{B}|2\rangle\,dp_{B} & = & \intop p_{B}^{n}\langle2|p_{B}\rangle\langle p_{B}|0\rangle\,dp_{B}\\
 & = & \intop x_{B}^{n}\langle0|x_{B}\rangle\langle x_{B}|2\rangle\,dx_{B}
\end{eqnarray*}
so that the second and third terms of equations  (\ref{eq:PnNL})
cancel. We obtain for $N=2$ that \textcolor{black}{$\Delta_{inf}^{2}(P_{b}^{2})=10.1351$.}\textcolor{red}{}

\textcolor{black}{Continuing for higher $N$, we obtain for $N=3$
$\Delta_{inf}^{2}(P_{b}^{3})=477.081$ and $\Delta_{inf}(P_{b}^{3})=21.8422$;
for $N=4$, $\Delta_{inf}^{2}(P_{b}^{4})=10982.8$ and $\Delta_{inf}(P_{b}^{4})=104.799$;
and for $N=5$, $\Delta_{inf}^{2}(P_{b}^{5})=795639$ and $\Delta_{inf}(P_{b}^{5})=891.986$.
Identical results are obtained for the inferred variances in $X^{N}$.}\textcolor{blue}{}

\section{Evaluation of $\left|\langle\left[n_{b},X_{b}^{N}\right]\rangle\right|_{inf}$
and $\left|\langle\left[n_{b},P_{b}^{N}\right]\rangle\right|_{inf}$}

The expressions for the terms $\left|\langle\left[n_{b},X_{b}^{N}\right]\rangle\right|_{inf}$
and $\left|\langle\left[n_{b},P_{b}^{N}\right]\rangle\right|_{inf}$
are calculated from Eq. (\ref{eq:C}) and the first line of Eq. (\ref{eq:eqncal2}).
Using that $X_{b}=b+b^{\dagger}$ and $P_{b}=(b-b^{\dagger})/i$,
we note that on evaluating the expectation value for the NOON states
given in Eq. (1), the only nonzero contributions involve terms of
the form $\left\langle b^{N}\right\rangle $ and $\left\langle b^{\dagger N}\right\rangle $:
\begin{eqnarray}
\left|\langle\left[n_{b},X_{b}^{N}\right]\rangle\right|_{inf} & = & N\left|-\langle b^{N}\rangle+\langle b^{\dagger N}\rangle\right|_{inf}\nonumber \\
\left|\langle\left[n_{b},P_{b}^{N}\right]\rangle\right|_{inf} & = & N\left|\langle b^{N}\rangle+(-1)^{N+1}\langle b^{\dagger N}\rangle\right|_{inf}\nonumber \\
\label{eq:RHSCriteriaXP-1}
\end{eqnarray}
\textcolor{red}{}\textcolor{green}{} We evaluate $\langle b^{N}\rangle=Tr\left[\rho_{red,x}b^{N}\right]$
and $\langle b^{\dagger N}\rangle=Tr\left[\rho_{red,x}\left(b^{\dagger}\right)^{N}\right]$
using the reduced density matrix given in Eq. (\ref{eq:redx}):\textcolor{black}{{}
}
\begin{eqnarray}
\langle\hat{b}^{N}\rangle_{inf,x} & = & \frac{e^{i\phi}}{2P\left(x\right)}\sqrt{N!}\langle x|0\rangle\langle x|N\rangle\label{eq:start}\\
\langle\left(\hat{b}^{\dagger}\right)^{N}\rangle_{inf,x} & = & \frac{e^{-i\phi}}{2P\left(x\right)}\sqrt{N!}\langle x|0\rangle\langle x|N\rangle\nonumber 
\end{eqnarray}
 On integrating over all possible values we get:
\[
\left|\langle\left[n_{b},X_{b}^{N}\right]\rangle\right|_{inf}=N\sqrt{N!}|\sin\phi|\intop_{-\infty}^{\infty}|\langle x|N\rangle\langle0|x\rangle|\,dx
\]
For $\left|\langle\left[n_{b},P_{b}^{N}\right]\rangle\right|_{inf}$
the expression for $N$ odd is given by:
\begin{eqnarray*}
\left|\langle\left[n_{b},P_{b}^{N}\right]\rangle\right|_{inf} & = & N\sqrt{N!}|\cos\phi|\intop_{-\infty}^{\infty}|\langle x|N\rangle\langle0|x\rangle|\,dx
\end{eqnarray*}
while for $N$ even we find:
\begin{eqnarray}
\left|\langle\left[n_{b},P_{b}^{N}\right]\rangle\right|_{inf} & = & N\sqrt{N!}|\sin\phi|\intop_{-\infty}^{\infty}|\langle x|N\rangle\langle0|x\rangle|\,dx\nonumber \\
\label{eq:end}
\end{eqnarray}
We obtain for $N=2$ that $\left|\langle\left[n_{b},P_{b}^{N}\right]\rangle\right|_{inf}=1.93577$
with $\phi=\pi/2$; for $N=3$, $\left|\langle\left[n_{b},P_{b}^{N}\right]\rangle\right|_{inf}=4.53$
with $\phi=0$; for $N=4$, $\left|\langle\left[n_{b},P_{b}^{N}\right]\rangle\right|_{inf}=11.2024$
with $\phi=\pi/2$; and for $N=5,$ $\left|\langle\left[n_{b},P_{b}^{N}\right]\rangle\right|_{inf}=29.5504$
with $\phi=0$. 

\section{Including losses\label{sec:SteeringCriteria_Losses}}

\textcolor{green}{}The detected fields $\hat{a}_{det}$, $\hat{b}_{det}$
are given by\textcolor{black}{{} 
\begin{eqnarray*}
a_{det} & = & \sqrt{\eta_{a}}a+\sqrt{1-\eta_{a}}a_{v}\\
a_{loss} & = & -\sqrt{1-\eta_{a}}a+\sqrt{\eta_{a}}a_{v}
\end{eqnarray*}
with similar definitions for the mode operators $b_{det}$ and $b_{loss}$.
}Using these transformations it is possible to write the operators
$a$, $b$ and hence the NOON state $|\psi\rangle$ of Eq. (\ref{eq:noon-state})
in terms of $a_{det}^{\dagger}$, $a_{loss}^{\dagger}$, $b_{det}^{\dagger}$
and $b_{loss}^{\dagger}$. \textbf{}We will denote the vacuum state
for all four modes by $|0\rangle$. The density operator $\rho=|\psi\rangle\langle\psi|$
can then also be expressed in terms of these operators.  Since
we are not interested in the modes $a_{loss}$ and $b_{loss}$ (which
we label $A,loss$ and $B,loss$) we take the trace over the states
of the loss mode to evaluate $\rho'\equiv Tr_{A,loss;B,loss}\rho$.
After using the binomial expansion for terms such as $\left(\sqrt{\eta_{a}}a_{det}^{\dagger}-\sqrt{(1-\eta_{a})}a_{loss}^{\dagger}\right)$
and performing the trace, the reduced density operator for the detected
modes is:\textcolor{blue}{}

\begin{widetext}

\begin{eqnarray}
\rho^{\prime} & = & \frac{1}{2}\left[\sum_{s}\begin{pmatrix}N\\
N-s
\end{pmatrix}\left(\eta_{a}\right)^{N-s}\left(1-\eta_{a}\right)^{s}\,|N-s\rangle_{A,det}\langle N-s|\otimes|0\rangle_{B,det}\langle0|+\left(\sqrt{\eta_{a}\eta_{b}}\right)^{N}e^{-i\phi}\,|N\rangle_{A,det}\langle0|\otimes|0\rangle_{B,det}\langle N|\right.\nonumber \\
 & + & \left.\left(\sqrt{\eta_{a}\eta_{b}}\right)^{N}e^{i\phi}\,|0\rangle_{A,det}\langle N|\otimes|N\rangle_{B,det}\langle0|+\sum_{s}\begin{pmatrix}N\\
N-s
\end{pmatrix}\left(\eta_{b}\right)^{N-s}\left(1-\eta_{b}\right)^{s}\,|0\rangle_{A,det}\langle0|\otimes|N-s\rangle_{B,det}\langle N-s|\right]\label{eq:density_operator_losses}
\end{eqnarray}

\end{widetext}

\subsection{Calculating $\Delta_{inf}^{2}(P_{b}^{N})$ and $\Delta_{inf}^{2}(X_{b}^{N})$}

The $\Delta_{inf}^{2}(P_{b}^{N})$ and $\Delta_{inf}^{2}(X_{b}^{N})$
are the inferred variances of quantities $P_{b}^{N}$ and $X_{b}^{N}$
due to a measurement in $X_{a}$. These are given by (\ref{eq:PinfN})
and (\ref{eq:XinfN}). We evaluate these inferred variances using
the density operator for modes $a_{det}$ and $b_{det}$ given in
Eq. (\ref{eq:density_operator_losses}). For the inferred variances
we evaluate the density operator $\rho''$, where we consider that
the mode $A,det$ is in the state $|x\rangle$. This density operator
is given by:\begin{widetext}
\begin{eqnarray*}
\rho'' & = & \frac{|x\rangle_{A,det}\langle x|\rho'|x\rangle_{A,det}\langle x|}{P\left(x\right)}\\
 & = & \frac{1}{2P\left(x\right)}\left[\sum_{s}\begin{pmatrix}N\\
N-s
\end{pmatrix}\left(\eta_{a}\right)^{N-s}\left(1-\eta_{a}\right)^{s}\langle x|N-s\rangle_{Ad}\langle N-s|x\rangle|x\rangle_{Ad}\langle x|\otimes|0\rangle_{Bd}\langle0|\right.\\
 &  & +\left(\sqrt{\eta_{a}\eta_{b}}\right)^{N}e^{-i\phi}\,\langle x|N\rangle_{Ad}\langle0|x\rangle|x\rangle_{Ad}\langle x|\otimes|0\rangle_{Bd}\langle N|+\left(\sqrt{\eta_{a}\eta_{b}}\right)^{N}e^{i\phi}\,\langle x|0\rangle_{Ad}\langle N|x\rangle|x\rangle_{Ad}\langle x|\otimes|N\rangle_{Bd}\langle0|\\
 &  & +\sum_{s}\begin{pmatrix}N\\
N-s
\end{pmatrix}\left(\eta_{b}\right)^{N-s}\left(1-\eta_{b}\right)^{s}\langle x|0\rangle_{Ad}\langle0|x\rangle|x\rangle_{Ad}\langle x|\otimes|N-s\rangle_{Bd}\langle N-s|\Biggl]
\end{eqnarray*}
\end{widetext}where  
\begin{eqnarray}
P\left(x\right) & = & Tr\left[|x\rangle_{A,det}\langle x|\rho'|x\rangle_{A,det}\langle x|\right]\nonumber \\
 & = & \frac{1}{2}\left[\sum_{s}\begin{pmatrix}N\\
N-s
\end{pmatrix}\left(\eta_{a}\right)^{N-s}\left(1-\eta_{a}\right)^{s}\,|\langle x|N-s\rangle|^{2}\right.\nonumber \\
 &  & +\sum_{s}\begin{pmatrix}N\\
N-s
\end{pmatrix}\left(\eta_{b}\right)^{N-s}\left(1-\eta_{b}\right)^{s}\,|\langle x|0\rangle|^{2}\Biggl]\label{eq:Px_losses}
\end{eqnarray}
Here we are using the following notation for the modes: $Ad\equiv A,det$
and $Bd\equiv B,det$. In order to compute $\Delta^{2}(P_{b}^{N}|x)$
and $\Delta^{2}(X_{b}^{N}|x)$, we trace out the $A,det$ mode to
get the reduced density operator for $B,det$ mode:\begin{widetext}
\begin{eqnarray}
\rho_{red,det,x} & = & Tr_{A,det}\left(\rho''\right)\nonumber \\
 & = & \frac{1}{2P\left(x\right)}\left[\sum_{s}\begin{pmatrix}N\\
N-s
\end{pmatrix}\left(\eta_{a}\right)^{N-s}\left(1-\eta_{a}\right)^{s}\langle x|N-s\rangle_{Ad}\langle N-s|x\rangle|0\rangle_{Bd}\langle0|\right.\nonumber \\
 &  & +\left(\sqrt{\eta_{a}\eta_{b}}\right)^{N}e^{-i\phi}\,\langle x|N\rangle_{Ad}\langle0|x\rangle|0\rangle_{Bd}\langle N|+\left(\sqrt{\eta_{a}\eta_{b}}\right)^{N}e^{i\phi}\,\langle x|0\rangle_{Ad}\langle N|x\rangle|N\rangle_{Bd}\langle0|\nonumber \\
 &  & +\sum_{s}\begin{pmatrix}N\\
N-s
\end{pmatrix}\left(\eta_{b}\right)^{N-s}\left(1-\eta_{b}\right)^{s}\left.\langle x|0\rangle_{Ad}\langle0|x\rangle|N-s\rangle_{Bd}\langle N-s|\right]\nonumber \\
\label{eq:densityop_redx}
\end{eqnarray}
\end{widetext}The inferred variances are defined as: 
\begin{eqnarray}
\Delta^{2}(X_{b}^{N}|x) & = & \langle\left(X_{b}^{N}\right)^{2}|x\rangle-\langle X_{b}^{N}|x\rangle^{2}\nonumber \\
\Delta^{2}(P_{b}^{N}|x) & = & \langle\left(P_{b}^{N}\right)^{2}|x\rangle-\langle P_{b}^{N}|x\rangle^{2}\label{eq:InfVariancesXP}
\end{eqnarray}

Next we evaluate $\langle X_{b}^{n}|x\rangle=Tr\left[\rho_{red,det,x}X_{b}^{n}\right]$
and $\langle P_{b}^{n}|x\rangle=Tr\left[\rho_{red,det,x}P_{b}^{n}\right]$
using the density operator given in Eq. (\ref{eq:densityop_redx})
obtaining:\begin{widetext}

\begin{eqnarray}
\langle X_{b}^{n}|x\rangle & = & \frac{1}{2P\left(x\right)}\left[\sum_{s}\begin{pmatrix}N\\
N-s
\end{pmatrix}\left(\eta_{a}\right)^{N-s}\left(1-\eta_{a}\right)^{s}\langle x|N-s\rangle_{Ad}\langle N-s|x\rangle\intop x_{B}^{n}\langle0|x_{B}\rangle\langle x_{B}|0\rangle\,dx_{B}\right.\nonumber \\
 & + & \left(\sqrt{\eta_{a}\eta_{b}}\right)^{N}e^{-i\phi}\langle x|N\rangle_{Ad}\langle0|x\rangle\intop x_{B}^{n}\langle N|x_{B}\rangle\langle x_{B}|0\rangle dx_{B}+\left(\sqrt{\eta_{a}\eta_{b}}\right)^{N}e^{i\phi}\langle x|0\rangle_{Ad}\langle N|x\rangle\intop x_{B}^{n}\langle0|x_{B}\rangle\langle x_{B}|N\rangle dx_{B}\nonumber \\
 & + & \left.\sum_{s}\begin{pmatrix}N\\
N-s
\end{pmatrix}\left(\eta_{b}\right)^{N-s}\left(1-\eta_{b}\right)^{s}\,\langle x|0\rangle_{Ad}\langle0|x\rangle\intop x_{B}^{n}\langle N-s|x_{B}\rangle\langle x_{B}|N-s\rangle\,dx_{B}\right]\label{eq:Xbn}
\end{eqnarray}

\end{widetext}

\begin{widetext}

\begin{eqnarray}
\langle P^{n}\rangle_{inf,x} & = & \frac{1}{2P\left(x\right)}\left[\sum_{s}\begin{pmatrix}N\\
N-s
\end{pmatrix}\left(\eta_{a}\right)^{N-s}\left(1-\eta_{a}\right)^{s}\langle x|N-s\rangle_{A,det}\langle N-s|x\rangle\intop p_{B}^{n}\langle0|p_{B}\rangle\langle p_{B}|0\rangle\,dp_{B}\right.\nonumber \\
 & + & \left(\sqrt{\eta_{a}\eta_{b}}\right)^{N}e^{-i\phi}\langle x|N\rangle_{Ad}\langle0|x\rangle\intop p_{B}^{n}\langle N|p_{B}\rangle\langle p_{B}|0\rangle dp_{B}+\left(\sqrt{\eta_{a}\eta_{b}}\right)^{N}e^{i\phi}\langle x|0\rangle_{Ad}\langle N|x\rangle\intop p_{B}^{n}\langle0|p_{B}\rangle\langle p_{B}|N\rangle\,dp_{B}\nonumber \\
 & + & \left.\sum_{s}\begin{pmatrix}N\\
N-s
\end{pmatrix}\left(\eta_{b}\right)^{N-s}\left(1-\eta_{b}\right)^{s}\langle x|0\rangle_{A,det}\langle0|x\rangle\intop p_{B}^{n}\langle N-s|p_{B}\rangle\langle p_{B}|N-s\rangle\,dp_{B}\right]\label{eq:Pbn}
\end{eqnarray}

\end{widetext}

The value of the corresponding variances for $\Delta^{2}(X_{b}^{N}|x)$
and $\Delta^{2}(P_{b}^{N}|x)$ of equations (\ref{eq:InfVariancesXP})
is evaluated using the expressions given in equations (\ref{eq:Xbn})
and (\ref{eq:Pbn}) considering $n=N$ or $n=2N$.

\subsection{Inferred variances $\Delta_{inf}^{2}(n_{b})$ including losses}

$\Delta_{inf}^{2}(n_{b})$ is the inferred variance of $n_{b}$ due
to a measurement in $n_{a}$. In order to evaluate this variance we
will consider that the outcome in $n_{a}$ is $m$. We define $P(m)$
as the probability for obtaining the result $m$ for $n_{a}$. Next,
we evaluate the reduced density operator $\rho_{m}$ for the modes
$A,det$ and $B,det$ given that the outcome is $m$: 
\begin{eqnarray*}
\rho_{m} & = & \frac{1}{P\left(m\right)}\left[|m\rangle_{Ad}\langle m|\rho'|m\rangle_{Ad}\langle m|\right]\\
 & = & \left[\begin{pmatrix}N\\
m
\end{pmatrix}\eta_{a}^{m}\left(1-\eta_{a}\right)^{N-m}|m\rangle_{Ad}\langle m|\otimes|0\rangle_{Bd}\langle0|\right.\\
\\
 &  & +\sum_{s}\begin{pmatrix}N\\
N-s
\end{pmatrix}\eta_{b}^{N-s}\left(1-\eta_{b}\right)^{s}\\
 &  & \times|0\rangle_{Ad}\langle0|\otimes|N-s\rangle_{Bd}\langle N-s|\Biggl]/(2P(m))
\end{eqnarray*}
where 
\begin{eqnarray}
P\left(m\right) & = & Tr\left[|m\rangle_{A,det}\langle m|\rho'|m\rangle_{A,det}\langle m|\right]\nonumber \\
 & = & \frac{1}{2}\begin{pmatrix}N\\
m
\end{pmatrix}\eta_{a}^{m}\left(1-\eta_{a}\right)^{N-m}+\frac{1}{2}\label{eq:Pna}
\end{eqnarray}
In order to write the last line we have used that $\sum_{s}^{N}\begin{pmatrix}N\\
N-s
\end{pmatrix}\eta_{b}^{N-s}\left(1-\eta_{b}\right)^{s}=1$. 

Next we evaluate $\langle n_{B}\rangle_{inf,m}=Tr\left[\rho_{m}n_{B}\right]$
and $\langle n_{B}^{2}\rangle_{inf,m}=Tr\left[\rho_{m}n_{B}^{2}\right]$
obtaining:
\begin{eqnarray*}
\langle n_{B}\rangle_{inf,m} & = & \frac{1}{2}\frac{\sum_{s}\begin{pmatrix}N\\
N-s
\end{pmatrix}\eta_{b}\left(1-\eta_{b}\right)^{s}\delta_{m,0}\left(N-s\right)}{P\left(n_{A}=m\right)}\\
\\
\langle n_{B}^{2}\rangle_{inf,m} & = & \frac{1}{2}\frac{\sum_{s}\begin{pmatrix}N\\
N-s
\end{pmatrix}\eta_{b}\left(1-\eta_{b}\right)^{s}\delta_{m,0}\left(N-s\right)^{2}}{P\left(n_{A}=m\right)}\\
\end{eqnarray*}
Since $n_{A}=m=0$ is the only non-zero contribution for the statistical
moments we obtain: 
\begin{eqnarray*}
\langle n_{B}\rangle_{inf,0} & = & \frac{1}{2}\frac{N\eta_{b}}{P\left(n_{A}=0\right)}\\
\langle n_{B}^{2}\rangle_{inf,0} & = & \frac{1}{2}\frac{\eta_{b}\left(N-N\eta_{b}+N^{2}\eta_{b}\right)}{P\left(n_{A}=0\right)}\\
P\left(n_{A}=0\right) & = & \frac{1}{2}\left(\left(1-\eta_{a}\right)^{N}+1\right)
\end{eqnarray*}
Using the above results we evaluate the inferred variance for $m=0$,
which we denote by $\Delta_{inf}^{2}n_{b,0}$: 
\begin{eqnarray*}
\Delta_{inf}^{2}n_{b,0} & = & \frac{\eta_{b}\left(N-N\eta_{b}\right)+N\eta_{b}\left(1-\eta_{a}\right)^{N}\left(1-\eta_{b}+N\eta_{b}\right)}{\left(\left(1-\eta_{a}\right)^{N}+1\right)^{2}}\\
\end{eqnarray*}
In order to evaluate the variance of the inferred value $n_{B}$,
we sum over all possible values of $m$ obtaining: 
\begin{eqnarray}
\Delta^{2}n_{inf} & = & \sum_{m}^{N}P\left(n_{A}=0\right)\Delta^{2}n_{inf,m=0}\nonumber \\
 & = & \frac{\eta_{b}\left(N-N\eta_{b}\right)+N\left(1-\eta_{a}\right)^{N}\left(\eta_{b}-\eta_{b}^{2}+N\eta_{b}^{2}\right)}{2\left(\left(1-\eta_{a}\right)^{N}+1\right)}\nonumber \\
\label{eq:Infvariance_n}
\end{eqnarray}

\subsection{Evaluation of $\left|\langle\left[n_{b},X_{b}^{N}\right]\rangle\right|_{inf}$
and $\left|\langle\left[n_{b},P_{b}^{N}\right]\rangle\right|_{inf}$}

Full evaluation of the terms $\left|\langle\left[n_{b},X_{b}^{N}\right]\rangle\right|_{inf}$
and $\left|\langle\left[n_{b},P_{b}^{N}\right]\rangle\right|_{inf}$
(given by Eqs. (\ref{eq:C}) and (\ref{eq:eqncal2})) reveals that
for the lossy system and for $N\leq5$:\textcolor{green}{{} }
\begin{eqnarray*}
\left|\langle\left[n_{b},X_{b}^{N}\right]\rangle\right|_{inf} & = & N\left|-\langle b^{N}\rangle+\langle b^{\dagger N}\rangle\right|_{inf}\\
\left|\langle\left[n_{b},P_{b}^{N}\right]\rangle\right|_{inf} & = & N\left|\langle b^{N}\rangle+(-1)^{N+1}\langle b^{\dagger N}\rangle\right|_{inf}
\end{eqnarray*}
\textcolor{red}{}\textcolor{green}{} We evaluate $\langle b^{N}\rangle=Tr\left[\rho_{red,det,x}b^{N}\right]$
and $\langle b^{\dagger N}\rangle=Tr\left[\rho_{red,det,x}\left(b^{\dagger}\right)^{N}\right]$
using the reduced density matrix given in Eq. (\ref{eq:densityop_redx})
and performed the corresponding trace, we obtain 
\begin{eqnarray*}
\langle b^{N}\rangle_{inf,x} & = & \frac{1}{2P\left(x\right)}\left(\sqrt{\eta_{a}\eta_{b}}\right)^{N}e^{i\phi}\langle x|N\rangle\langle0|x\rangle\sqrt{N!}
\end{eqnarray*}
and $ $$\langle\left(b^{\dagger}\right)^{N}\rangle_{inf,x}=(\langle b^{N}\rangle_{inf,x})^{*}$.
Thus the expressions obtained are identical to (\ref{eq:start}-\ref{eq:end})
but replacing $\sqrt{N!}$ with $C_{\eta}=\sqrt{N!}\left(\sqrt{\eta_{a}\eta_{b}}\right)^{N}$.

\section{Proof of $N$th order quantum coherence and entanglement}

The question is how to prove experimentally that the system is indeed
in a superposition of the two number states $|N\rangle|0\rangle$
and $|0\rangle|N\rangle$ that are distinct by $N$ quanta in each
mode. For such a state, the density matrix $\rho$ has a nonzero off\textendash diagonal
element: 
\begin{equation}
\langle0|\langle N|\rho|0\rangle|N\rangle\neq0\label{eq:ncoh}
\end{equation}
We refer to the nonzero term (\ref{eq:ncoh}) as an \emph{$N$th order
quantum coherence}. The presence of this term distinguishes the superposition
of the two states $|N\rangle|0\rangle$, $|0\rangle|N\rangle$ from
a classical mixture of the two states. 

In the ideal scenario, the experiment generates only outcomes $0$
or $N$ for the number measurements $n_{a}$ or $n_{b}$. It is then
straightforward to show that any violation of an EPR steering inequality
is also a signature of an $N$th order quantum coherence (\ref{eq:ncoh}).
The objective is to construct the density operator $\rho$ for the
system and to prove that necessarily (\ref{eq:ncoh}) holds. Since
there are only two outcomes for each mode, any viable two-mode density
operator could be written in terms of four basis states $|0\rangle|0\rangle,|0\rangle|N\rangle,|N\rangle|0\rangle,|N\rangle|N\rangle$.
Supposing an EPR steering inequality to be violated, this will negate
the LHS model given by (\ref{eq:lhs}) and therefore also any fully
separable quantum model 
\begin{equation}
\rho=\sum_{R}P_{R}\rho_{a}\rho_{b}\label{eq:mixden-1}
\end{equation}
where $\rho_{a}$ and $\rho_{b}$ are density matrices for the single
modes $a$ and $b$ \cite{steer-1}. Thus the system cannot be in
any mixture of the basis states $|0\rangle|0\rangle,|0\rangle|N\rangle,|N\rangle|0\rangle,|N\rangle|N\rangle$
which are separable states. There are only certain remaining possibilities
for $\rho$ and these require non-zero off-diagonal elements. For
example, for the NOON state the results for number measurements $\hat{n}$
would be either $0$ or $N$ in one mode, correlated with $N$ or
$0$ in the other mode. Assuming that there is a nonzero probability
for the outcome for $|0\rangle|N\rangle$ and $|N\rangle|0\rangle$,
this ensures that the off-diagonal term $\langle0|\langle N|\rho|0\rangle|N\rangle$
is nonzero. Also, the failure of the separable model (\ref{eq:mixden-1})
ensures the system cannot be in the product state $(|N\rangle+|0\rangle)(|N\rangle+|0\rangle)/2$.
This implies the system is the entangled superposition of states $|0\rangle|N\rangle$
and $|N\rangle|0\rangle$.

\section{Proof of inequality (\ref{eq:uncertwomode-1})}

\textbf{}First we prove the the uncertainty relation $ $
\begin{equation}
(\Delta\hat{n}_{b})^{2}(\Delta_{inf}\hat{P}_{b}^{N})^{2}\geq\frac{1}{4}|\langle\hat{C_{b}}\rangle|_{inf}^{2}\label{eq:uncerappendix}
\end{equation}
which holds for any two-mode state. We follow the methods used in
Refs. \cite{ericpramix}. The variance is defined as $(\Delta\hat{n}_{b})^{2}=\sum_{n_{b}}P(n_{b})(n_{b}-\langle n_{b}\rangle)^{2}$
(denoting the outcomes of $\hat{n}_{b}$ by $n_{b}$). We can consider
marginals and joint distributions for the measurements on both modes
$a$ and $b$. Thus we write $(\Delta\hat{n}_{b})^{2}=\sum_{n_{b},p_{a}}P(n_{b},p_{a})(n_{b}-\langle n_{b}\rangle)^{2}$
and then $(\Delta\hat{n}_{b})^{2}=\sum_{n_{b},p_{a}}P(n_{b}|p_{a})P(p_{a})(n_{b}-\langle n_{b}\rangle)^{2}$.
Thus 
\begin{eqnarray*}
(\Delta\hat{n}_{b})^{2} & = & \sum_{p_{a}}P(p_{a})\sum_{n_{b}}P(n_{b}|p_{a})(n_{b}-\langle n_{b}\rangle)^{2}\\
 & \geq & \sum_{p_{a}}P(p_{a})\sum_{n_{a}}P(n_{b}|p_{a})(n_{b}-\langle n_{b}\rangle_{p_{a}})^{2}\\
 & = & \sum_{p_{a}}P(p_{a})(\Delta(n_{b}|p_{a}))^{2}
\end{eqnarray*}
where $(\Delta(n_{b}|p_{a}))^{2}=\sum_{n_{a}}P(n_{b}|p_{a})(n_{b}-\langle n_{b}\rangle_{p_{a}})^{2}$
and $\langle n_{b}\rangle_{p_{a}}$ is the mean of the conditional
distribution $P(n_{b}|p_{a})$. For each $p_{a}$ we have defined
the distribution $P(n_{b}|p_{a})$ as $P_{p_{a}}(n_{b})$ and we see
that the quantity $\sum_{n_{b}}P_{p_{a}}(n_{b})(n_{b}-X)^{2}$ where
$X$ is any constant, is minimised by the choice $X=\langle n_{b}\rangle_{p_{a}}=\sum_{n_{b}}P_{p_{a}}(n_{b})n_{b}$.
Next, we write 

\begin{eqnarray*}
(\Delta\hat{n}_{b})^{2}(\Delta_{inf}\hat{P}_{b}^{N})^{2} & \geq & \{\sum_{p_{a}}P(p_{a})(\Delta(n_{b}|p_{a}))^{2}\}\\
 &  & \ \ \ \ \{\sum_{p_{a}}P(p_{a})(\Delta(P_{b}^{N}|p_{a})^{2}\}\\
 & \geq & |\sum_{p_{a}}P(p_{a})\Delta(n_{b}|p_{a})\Delta(P_{b}^{N}|p_{a})|^{2}\\
 & \geq & \frac{1}{4}|\sum_{p_{a}}P(p_{a})|\langle C_{b}\rangle_{p_{a}}||^{2}=\frac{1}{4}|\langle\hat{C_{b}}\rangle|_{inf}^{2}
\end{eqnarray*}
where we apply the Cauchy-Schwarz inequality and use the uncertainty
relation (2) that holds for the state of $b$ conditioned on the measurement
result $p_{a}$ of mode $a$. This proves (\ref{eq:uncerappendix}).
Then we can say that for the mixture $\rho^{ab}$ of (\ref{eq:twomodemix})
(using that for a mixture it is true that \cite{take} $(\Delta O)_{\rho}^{2}\geq\sum_{i}P_{i}(\Delta O)_{i}^{2}$
where $O$ is any quantum observable, and also true that $(\Delta_{inf}P_{b}^{N})^{2}\geq\sum_{i}P_{i}(\Delta_{inf}P_{b}^{N})_{i}^{2}$
\cite{mallon}): 
\begin{eqnarray}
\{\sum_{i}P_{i}(\Delta\hat{n}_{b})_{i}^{2}\}(\Delta_{inf}\hat{P}_{b}^{N})^{2} & \geq & \{\sum_{i}P_{i}(\Delta n_{b})_{i}^{2}\}\nonumber \\
 &  & \ \ \ \ \{\sum_{i}P_{i}(\Delta_{inf}P_{b}^{N})_{i}^{2}\}\nonumber \\
 & \geq & |\sum_{i}P_{i}(\Delta n_{b})_{i}(\Delta_{inf}P_{b}^{N})_{i}|^{2}\nonumber \\
 & \geq & \frac{1}{4}|\sum_{i}P_{i}|\langle\hat{C_{b}}\rangle|_{inf,i}|^{2}\label{eq:result}
\end{eqnarray}
where we use the Cauchy-Schwarz inequality and that the uncertainty
relation (2) holds for each $\rho_{i}^{ab}$. Now we see that $\sum_{p_{a}}P(p_{a})|\langle C_{b}\rangle_{p_{a}}|=\sum_{p_{a}}P(p_{a})|\sum_{c_{b}}C_{b}P(C_{b}|p_{a})|$.
If the system is described by the mixture $\rho^{ab}$ then 
\begin{eqnarray}
\langle C_{b}\rangle_{p_{a}} & = & \sum_{c_{b}}C_{b}P(C_{b}|p_{a})\nonumber \\
 & = & \sum_{C_{b}}C_{b}\frac{P(C_{b},p_{a})}{P(p_{a})}\nonumber \\
 & = & \sum_{C_{b}}C_{b}\sum_{i}P_{i}\frac{P_{i}(C_{b},p_{a})}{P(p_{a})}\nonumber \\
 & = & \sum_{i}P_{i}\sum_{C_{b}}C_{b}\frac{P_{i}(p_{a})}{P(p_{a})}P_{i}(C_{b}|p_{a})\label{eq:eqn4}
\end{eqnarray}
where the subscript $i$ denotes the probabilities for the component
$\rho_{i}^{ab}$. We can write
\begin{eqnarray*}
|\sum_{i}P_{i}\sum_{C_{b}}C_{b}\frac{P_{i}(p_{a})}{P(p_{a})}P_{i}(C_{b}|p_{a})| & \leq & \sum_{i}P_{i}\frac{P_{i}(p_{a})}{P(p_{a})}\\
 &  & \ \ \times|\sum_{C_{b}}C_{b}P_{i}(C_{b}|p_{a})|
\end{eqnarray*}
Thus from (\ref{eq:eqn4}) 
\begin{eqnarray*}
\sum_{p_{a}}P(p_{a})|\langle C_{b}\rangle_{p_{a}}| & \leq & \sum_{i}P_{i}\sum_{p_{a}}P_{i}(p_{a})|\sum_{C_{b}}C_{b}P_{i}(C_{b}|p_{a})|\\
 & = & \sum_{i}P_{i}|\langle\hat{C_{b}}\rangle|_{inf,i}
\end{eqnarray*}
where $|\langle\hat{C_{b}}\rangle|_{inf,i}=\sum_{p_{a}}P_{i}(p_{a})|\sum_{C_{b}}C_{b}P_{i}(C_{b}|p_{a})|$.
Thus we have proved that $\sum_{i}P_{i}|\langle\hat{C_{b}}\rangle|_{inf,i}\geq\sum_{p_{a}}P(p_{a})|\langle C_{b}\rangle_{p_{a}}|$.
Hence we can write from (\ref{eq:result}): 
\begin{eqnarray*}
\{\sum_{i}P_{i}(\Delta n_{b})_{i}^{2}\}(\Delta_{inf}P_{b}^{N})^{2} & \geq & \frac{1}{4}|\sum_{i}P_{i}|\langle\hat{C_{b}}\rangle|_{inf,i}|^{2}\\
 & \geq & \frac{1}{4}|\sum_{p_{a}}P(p_{a})|\langle C_{b}\rangle_{p_{a}}||^{2}\\
 & = & \frac{1}{4}|\langle\hat{C_{b}}\rangle|_{inf}|^{2}
\end{eqnarray*}
This proves the inequality (\ref{eq:uncertwomode-1}). $\square$

\end{document}